\documentclass[twocolumn]{aastex62}
\usepackage{graphicx}
\usepackage{mathtools, amsmath}
\usepackage{longtable}
\usepackage{url}
\usepackage{color}
\usepackage[utf8]{inputenc}
\usepackage[titletoc,title]{appendix}

\shorttitle{Constraining stellar parameters and atmospheric dynamics of V~Oph}
\shortauthors{Rau et al.}

\begin{document}


\title{Constraining stellar parameters and atmospheric dynamics of the carbon AGB star V~Oph}


\author{Gioia Rau}
\affil{NASA/GSFC Code 667, Goddard Space Flight Center, Greenbelt, MD 20071, USA}
\affil{CRESST II - Department of Physics, Catholic University of America, Washington, DC 20064, USA} 
\affil{Department of Astrophysics, University of Vienna, T\"{u}kenschanzstrasse 17, 1180 Vienna, Austria}
\email{gioia.rau@nasa.gov}

\author{Keiichi Ohnaka}
\affil{Universidad Catolica del Norte, Instituto de Astronomia, Avenida Angamos 0610, Antofagasta, Chile}

\author{Markus Wittkowski}
\affil{European Southern Observatory (ESO), Karl-Schwarzschild-Str. 2, 85748 Garching bei M\"{u}nchen, Germany}

\author{Vladimir Airapetian}
\affil{NASA/GSFC Code 667, Goddard Space Flight Center, Greenbelt, MD 20071, USA}

\author{Kenneth G.~Carpenter}
\affil{NASA/GSFC Code 667, Goddard Space Flight Center, Greenbelt, MD 20071, USA}

\begin{abstract}

Molecules and dust produced by the atmospheres of cool evolved stars contribute to a significant amount of the total material found in the interstellar medium. To understand the mechanism behind the mass loss of these stars, it is of pivotal importance to investigate the structure and dynamics of their atmospheres.

Our goal is to verify if the extended molecular and dust layers of the carbon-rich asymptotic giant branch (AGB) star V~Oph, and their time variations, can be explained by dust-driven winds triggered by stellar pulsation alone, or if other mechanisms are operating.

We model V~Oph mid-infrared interferometric VLTI-MIDI data ($8$--$13~\mu$m), at phases $0.18$, $0.49$, $0.65$, together with literature photometric data, using the latest-generation self-consistent dynamic atmosphere models for carbon-rich stars: DARWIN.

We determine the fundamental stellar parameters: $T_\text{eff} = 2600~$K, $L_\text{bol} = 3585~$L$_{\odot}$, $M = 1.5~$M$_{\odot}$, $C/O = 1.35$, $\dot{M} = 2.50\cdot10^{-6}$M$_{\odot}$/yr. We calculate the stellar photospheric radii at the three phases: $479$, $494$, $448$~R$_{\odot}$; and the dust radii: $780$, $853$, $787$~R$_{\odot}$. The dynamic models can fairly explain the observed $N$-band visibility and spectra, although there is some discrepancy between the data and the models, which is discussed in the text.

We discuss the possible causes of the temporal variations of the outer atmosphere, deriving an estimate of the magnetic field strength, and computing upper limits for the Alfv\'{e}n waves velocity. In addition, using period-luminosity sequences, and interferometric modeling, we suggest V~Oph as a candidate to be reclassified as a semi-regular star.

\end{abstract}

\keywords{stars: winds, stars: mass-loss, stars: AGB, techniques: interferometry, stars: carbon, stars: circumstellar matter}

\section{Introduction} \label{intro}
After moving off the main sequence (MS), stars that are born with an initial mass in the range between~$0.8$~and~$\sim8$~M$_{\odot}$ evolve to become red giant (RG) and horizontal branch (HB) stars, and then move forward ascending the asymptotic giant branch (AGB).


After experiencing several third dredge-ups, the mechanism responsible for transforming AGB stars from O-rich to C-rich \citep{ibenrenzini83}, AGB stars can change their spectral appearance and become carbon stars. Being surrounded by dust and gas, AGB stars are one of the most important contributors to the enrichment of the interstellar medium. In particular, the contribution from carbon-rich AGB stars is essential for producing carbon-bearing molecules and dust, in particular C$_2$, C$_3$, C$_2$H$_2$, CN, and HCN molecules (see e.g., \citealp{olofsson93, cernicharo00, gong15}); and amorphous carbon (amC) and silicon carbide (SiC) dust (see e.g., \citealp{yamamura00, loidl01, bladh19}). 

As the star evolves, and becomes bigger in dimension, brighter, and cooler, it will eventually begin to pulsate. The pulsation will generate shock waves, and, together with convection, lead to strongly extended molecular
atmospheres. There, if the conditions are favorable (i.e., temperature and density sufficiently low and high, respectively), dust could eventually form. Up to now, the commonly accepted scenario for the mass loss of carbon stars is the following: if the opacity of the amC dust is high enough, the radiation pressure acting upon the dust grains will eventually provide enough momentum to the grains to accelerate them, and drag along by collision the gas, driving an outflow (stellar wind) from the star. (see e.g., \citealp{Fleischer92, HoefnerDorfi}).

\cite{hofner03} modeled this scenario, solving the coupled equations of hydrodynamics, with frequency-dependent radiative transfer and time-dependent formation, growth, and evaporation of dust grains. The DARWIN models are the results of these dynamic atmospheres calculations. These models have successfully reproduced observations of carbon-rich stars, e.g.~line profile variations \citep{nowotny10} and time-dependent spectroscopic data \citep{Gautschy-Loidl04, nowotny13}. However, several discrepancies remain when comparing those with interferometric observations (e.g.~\citealp{sacuto11, cruzalbes13, klotz13, vanbelle13, rau15, rau16, rau17, wittkowski18}).

If this means that pulsation and radiation pressure alone are not able to explain the amount of mass lost by these stars, other mechanisms could play a role in helping driving the winds of cool giant stars. The possibility of (Magnetic) Alfv\'{e}n waves driving stellar winds and producing clumpy mass loss is discussed in e.g.~\cite{woitke06, vidotto06, hofner07, hoefnerandolofsson18, rau19}, and very recently by \cite{yasuda19}.

\object{V~Oph} is currently classified as a carbon-rich, Mira star (e.g., \citealp{asas}), with a period of $297~$days. Its distance estimation from \cite{gaiaparallaxes} is $731\pm22~$pc, and the variability amplitude is $4.3~$mag. Table~\ref{table_starsparam} shows this and other parameters of the star.

\cite{ohnaka07} (hereafter OH07) modeled the temporal variation of the physical properties of the outer atmosphere and circumstellar dust shell of V~Oph, based on the mid-infrared interferometric data taken at $8$--$13~\mu$m with the MID-infrared Interferometric instrument (MIDI, \citealp{midi2003}) observations at the Very Large Telescope Interferometer (VLTI). They derived the physical properties of the molecular outer atmosphere of C$_2$H$_2$ and HCN and the inner dust envelope based on semi-empirical models. However, as the same author suggests, to better understand the physical processes responsible for molecule and dust formation close to the star, a comparison of the observations with self-consistent dynamical models is necessary. Such a comparison has not yet been attempted, and is not presented in OH07.

Our intent is thus to analyze the previously existing data on the carbon-rich AGB star V~Oph (reported in \citealp{ohnaka07}), modeling them with the state-of-the-art grid of dynamic atmosphere models for C-rich stars from \cite{mattsson10} and \cite{erik14}. We aim also to explore and untangle the possible explanations (see also Section~$5$ in OH07) on whether the prominent role in causing C$_2$H$_2$ molecular extended layers is due to the dust-driven winds triggered by large-amplitude stellar pulsation alone, or if other physical mechanisms are a stake in V~Oph, such as e.g.~winds driven by Alfv\'{e}n waves (e.g., \citep{airapetian00, airapetian10, airapetian15}). We calculate the Alfv\'{e}n wave speed to investigate this possible mechanism, which has not yet been tried before.

We summarize and describe the archive observations of the C-rich AGB stars V~Oph in Sect.~\ref{data}, together with its basic parameters. Sect.~\ref{DARWIN models} illustrates the comparison of the observables with the self-consistent dynamic atmosphere models used. Sect.~\ref{results} presents our results, which will be discussed in Sect.~\ref{discussion}, including a comparison with the evolutionary tracks, a speculative alternative scenario for the mass loss, and a possible reclassification of V~Oph to semi-regular star. We conclude in Sect.~\ref{conclu} with perspectives for future works.

\begin{table*}[!htbp]
\caption{\label{table_starsparam} Main parameters of V~Oph.}
\centering
\begin{tabular}{llllllllll}
\hline
\hline
 Name  & Variability  &   $P$~\textsuperscript{a} & $\Delta{V}$~\textsuperscript{a}  & $d$~\textsuperscript{b}  & $L_\text{bol}$~\textsuperscript{c}    &  $\dot{M}$~\textsuperscript{d} &  $\dot{M}$~\textsuperscript{e}  &  $\dot{M}$~\textsuperscript{f} & $v_\text{e}~\textsuperscript{g}$\\
 &  Type~\textsuperscript{a} &   [d] &   [mag] & [pc] & [L$_{\odot}$]  &   [$10^{-7}$~M$_{\odot}$/yr] &   [$10^{-7}$~M$_{\odot}$/yr]  & [$10^{-7}$~M$_{\odot}$/yr]   & [km~s$^{-1}$]\\
\hline
V~Oph   &   M & 297 &  4.3 & $ 731 \pm 22$  & $3585$  &  $1.2\pm\ldots$ & $1.4\pm\ldots$  & $0.5\pm\ldots$ & $\sim7.5$\\
\hline
\hline
\end{tabular}\\
\textbf{Notes}. The ``\ldots'' indicate that no literature value is given. (a): \cite{GCVS}. (b): GAIA DR2: \cite{gaia2016, gaia2018, gaiaparallaxes}. (c): $L_\text{bol}$ is the bolometric luminosity derived from the SED fitting. (d) \cite{whitelock06}. (e) \cite{bergeat05}. (f) \cite{groenewegen99}. (g) observed terminal gas expansion velocity from CO lines \citep{groenewegen99}.
\end{table*}

\section{Observational data}\label{data}

\object{V~Oph} has been classified up to now as a Mira variable star. This target has been observed by OH07 with the Very Large Telescope Interferometer (VLTI) of ESO Paranal Observatory with the mid-infrared interferometric recombiner (MIDI, \citealp{midi2003}) using the $8.2~$m Unit Telescopes. MIDI, now decommissioned, provided wavelength-dependent visibilities, photometry and differential phases in the $N$-band, i.e.~from $8$ to $13$~$\mu$m. 

OH07 present the wavelength-dependent visibilities, differential phases, and spectra taken from $8$ to $13~\mu$m. For the journal of observations we refer to OH07, their Table~$1$. The star, observed over $6$ nights at different periods, shows evidence of interferometric variability, as reported by the same authors. For the modeling described in Sect.~\ref{DARWIN models} we adopted the dataset binning as in OH07, i.e.~a binning into three epochs: $\phi = 0.18$, $\phi = 0.49$, and $\phi = 0.65$.

The main parameters of the star, namely variability class, period, amplitude of variability, distance, and mass-loss rate, are shown in Table~\ref{table_starsparam}.

\begin{figure*}
\centering
\includegraphics[bb=65 62 533 699, clip=true, scale=0.7, angle=90]{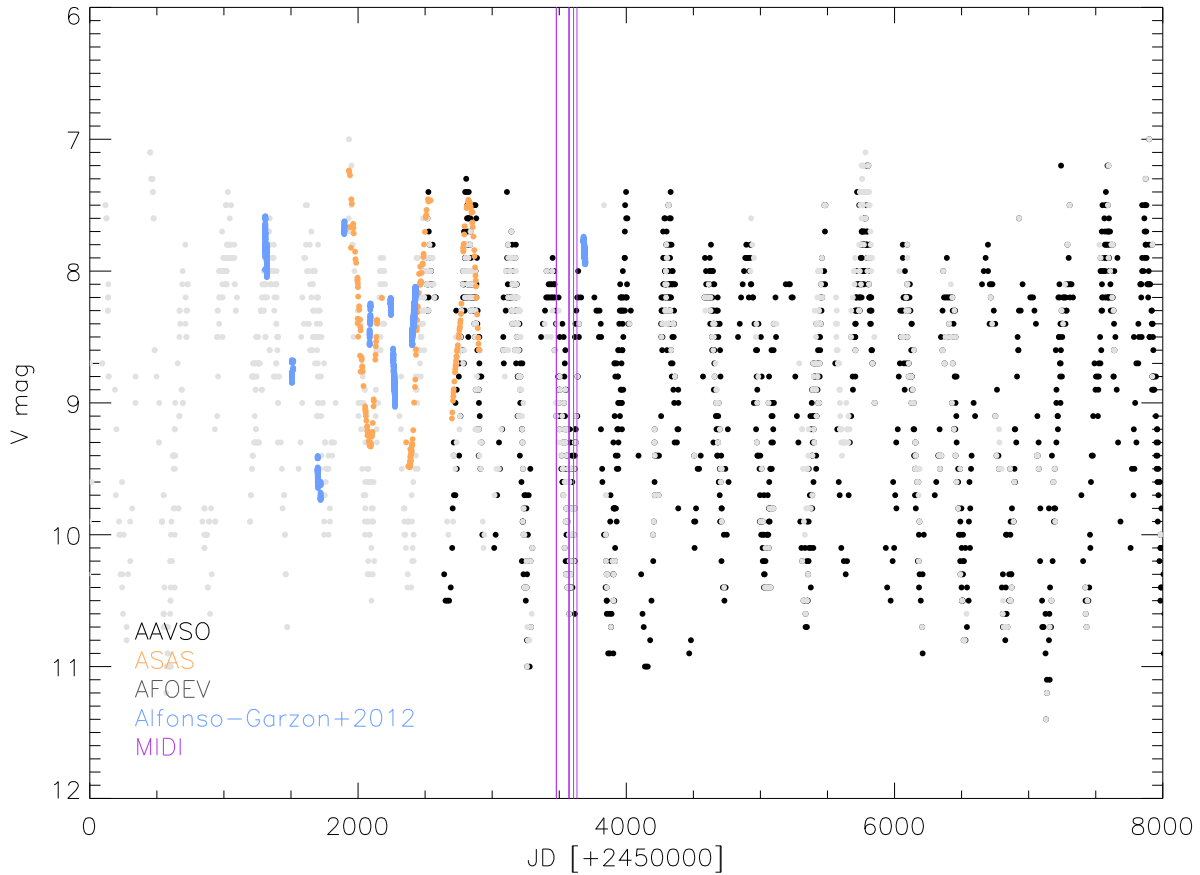}
\caption{AAVSO, ASAS, AFOEV, and Alfonso-Garzon light curves of V~Oph, in black, orange, grey, and light blue respectively. The six vertical purple lines denote the epochs of the MIDI observations.}
\label{newlight}
\end{figure*}
\begin{figure}
\center{
\includegraphics[bb=65 62 533 699, scale=0.38, clip=true, angle=90]{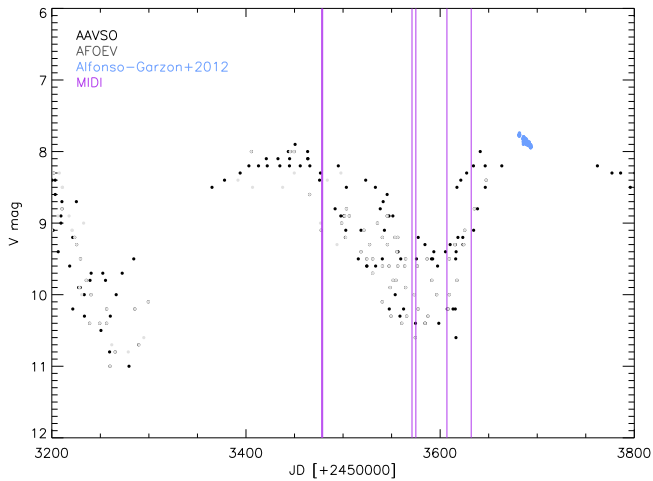}
\caption{As Fig.~\ref{newlight}, but with a blow-up around the MIDI observations.}}
\label{newlight_blowup}
\end{figure}

V~Oph visual light curves were collected from archive data of \cite{aavso, asas, afoev, alf12}, and showed in Fig.~\ref{newlight}, where the time of the MIDI observation is marked in purple, and some light curve irregularities are noticeable. 
The nature of this irregularities are discussed in Sect.~\ref{disc_sedvis}. A blow-up around the MIDI observations is show in Fig.~\ref{newlight_blowup}.

\subsection{Photometry}\label{phot}

We collected photometry from the literature in the $V$, $R$,  $L$, and $M$ photometric bands \citep{whitelock06, dirbe, asas, denis, 2mass, iras-satellite}. A mean value was calculated for each filter, with amplitudes derived from the variability. For the other bands where no light curves were available ($B$, $I$, $J$, $H$, $K$), we collected values from the literature and averaged them. The standard deviations have been adopted as the errors on those values. In the case of bands with only one value and no associated error from the literature, we assumed a conservative $20$\% of the value as error.

\section{Modeling with dynamic atmosphere models}\label{DARWIN models}

\subsection{Models description}\label{modeldescription}

We compared the observations with self-consistent dynamic atmosphere models for carbon-rich evolved stars. Using a grid of models representing a range of stellar parameters \citep{mattsson10, erik14}, we calculated synthetic photometry and synthetic interferometric profiles. Those were then compared with the observed (literature) photometric and interferometric data. The dynamic atmosphere models give varying results from cycle to cycle; therefore we compared the data with each time-step. The details on the fitting procedure are given in Sect.~\ref{fitting_procedure}).
 
Those models result from solving the system of equations for hydrodynamics and spherically symmetric frequency-dependent radiative transfer, plus equations describing the time-dependent dust formation, growth, and evaporation. The models start with an initial hydrostatic structure; then a ``piston'' simulates the pulsation of the star, varying the inner boundary below the stellar photosphere. In C-rich stars amorphous carbon (amC) and silicon carbide (SiC) are the most abundant dust species found (see later). We underline that in those models only the dust formation of amorphous carbon is calculated in a self-consistent way, through the ``method of moments'' \citep{gauger90,gail88}. We refer to \cite{HoefnerDorfi}, \cite{hofner99}, \cite{hofner03}, \cite{hofner16}, \cite{bladh19b} for a detailed description of the modeling approach; the applications to observations are described in \cite{loidl99}, \cite{Gautschy-Loidl04}, \cite{nowotny10}, \cite{walter2}, \cite{sacuto11}, \cite{rau15}, \cite{rau16}, \cite{rau17}. 
 
These models are characterized by a series of parameters, such as effective temperature $T_{\text{eff}}$, luminosity $L$, mass $M$, carbon-to-oxygen ratio $C/O$, piston velocity amplitude $\Delta_{\text{u}}$, and the parameter $f_\text{L}$ used in the calculations to adjust the luminosity amplitude of the model. From the hydrodynamic calculations the mean degree of condensation, wind velocity, and the mass-loss rate are calculated. Each model is described by a set of ``time-steps'', i.e.~snapshots, each at different phases of the stellar pulsation.

The COMA code \citep{aringer00, aringer09}, and the subsequent radiative transfer, was used to calculate the synthetic spectra, intensity profiles, and visibilities, from the temperature and pressure stratification of the dynamical models at each time step. To derive the synthetic photometry we integrated the synthetic spectra over the selected filters mentioned in Sect.~\ref{phot}. The abundances of the atomic, molecular, and dust species were calculated, from the temperature-density structure vs.~radius, considering the equilibrium for ionization and molecule formation. Assuming local thermal equilibrium (LTE), the atomic and molecular spectral line strengths were computed, together with the continuous gas opacity. The molecular opacities data are listed in \cite{cristallo07}, while \cite{aringer09} shows the molecular spectral line strengths, and are consistent with the data used for constructing the models. 

From the output of the dynamic atmosphere models, which are subsequently used in the COMA modeling, we took the amount of dust condensed into amorphous carbon (amC), in g/cm$^3$. We treated the amC dust opacity consistently (\citealp{roleau_and_martin} in small particle limit (SPL) - see also \citealp{erik14} for further details on the dust treatment), since amC, as opposed to SiC, is considered in the models so there has been no need to add it a posteriori with COMA. On the other hand, SiC is added artificially a posteriori, with COMA.

Following \cite{sacuto11, rau15, rau17}, the percentage of condensed material is as follows: $90$~\% amorphous carbon, using data from \cite{roleau_and_martin}, and $10$\% silicon carbide, based on \cite{pegourie}. 

\subsection{The fitting procedure}\label{fitting_procedure}


We follow the fitting procedure approach described extensively in \cite{rau17}. \textit{First} we compared the photometric observations taken from the literature to the whole grid of $540$ dynamic atmosphere models. The parameter space of the grid can be found in \cite{erik14} (i.e.~their Figure~$2$). 

We have calculated the model best fitting the photometric literature data among the whole grid, with the method of least squares ($\chi^2$) as follows: 

\begin{equation}\label{chisq}
\chi^2 = \sum_{(k=0, j=o)}^{N} \frac{[m_{\text{MOD}}(k,j) - m_{\text{OBS}}(j)]^2}{\sigma_{\text{OBS}}^2 (j)},
\end{equation}

where m$_{\text{MOD}}$ is the model magnitude and m$_{\text{OBS}}$ is the observed magnitude. The magnitudes have been normalized to the $K$-band magnitude, where the variability is the smallest, i.e.~$m = m^{\text{K-band}} - m^{\text{band}}$, and this is the quantity plotted in the ordinate in Fig.~\ref{phot-voph}. $\sigma_{\text{OBS}}$ is the error on the photometry. The parameters of this model, together with its corresponding best-fit photometric time-step, are listed in Table~\ref{tab_param_dyn}.

\begin{table*}
\centering
\caption{\label{tab_param_dyn} Summary of the best-fit model for each type of observation: photometry, and interferometry. Listed are the corresponding values of the $\chi^2$, and the parameters of the models.}
\begin{tabular}{llllllllllllllll}
\hline
\hline
 & $T_{\text{eff}} $ & $\log(L)$ & $M$ & $P$ & $\log~g$ & $C/O$ & $\Delta u_{\text{p}}$ & $f_{\text{L}}$ & $\dot{M}$ & $v_\text{e}$~\textsuperscript{a} & $\lambda_{\text{fit range}}$ & ${\chi^2}_{\text{red}}$ & $\phi_{\text{observ}}$  & $\phi_{\text{mod}}$  \\
 &$[K]$ & [L$_{\odot}$] & [M$_{\odot}$] & [d] & [cm$^2$~s$^{-1}$]  &  & [km~s$^{-1}$] &   & [10$^{-6}$M$_{\odot}$~yr$^{-1}$] & [km~s$^{-1}$] &  [$\mu$m]  &   & & & \\
\hline
\bf{V~Oph}  &                   &            &   &   &    &  &           &       &    &  &  \\
\hline 
Photom & 2600&  4.00&   1.5&525&        -0.79& 1.35&      6&      1& 2.51 & $\sim8$  & [0.4-25]&  \hphantom{0}1.72 &\ldots & \ldots \\
NO $\dot{M}$ & 2600&    4.00&   2.00 &  525 &  -0.66  &  1.35 &          6&       1&  \ldots & $\ldots$  & [0.4-25]&  \hphantom{0}1.53 & \ldots & \ldots \\
\hline
Interf$\phi_{1}$ & 2600&  4.00&   1.5& 525 &      -0.79& 1.35&      6&      1&  2.51 &  $\sim8$  &  [8-13]   &   \hphantom{0}1.36  &  0.18 & 0.85 \\
Interf$\phi_{2}$ & 2600&  4.00&   1.5& 525 &      -0.79& 1.35&      6&      1&  2.51 &  $\sim8$  &  [8-13]   &   \hphantom{0}5.03  &  0.49 & 0.10\\
Interf$\phi_{3}$ & 2600&  4.00&   1.5& 525 &      -0.79& 1.35&      6&      1&  2.51 &  $\sim8$  &  [8-13]   &   \hphantom{0}1.18  &  0.65 & 0.75\\
\hline 
\hline
\end{tabular}\\
\textbf{Notes}. (a) Model gas velocity from online material at \cite{erik14}
\end{table*}

\textit{Second}, from the computed intensity distribution, we produced the synthetic visibilities, following the approach of \cite{davis00} and \cite{tangoanddavis02}, for all the time-steps belonging to the best-fitting model for photometry. We calculated the Hankel transformation of the intensity distribution $I$, and then compared them to V~Oph interferometric MIDI data, determining the best-fitting model using a chi-square analysis as from Eq.~\ref{chisq}, where visibilities (and their errors) replace the magnitudes (and their errors). The error on the interferometric visibilities is of the order of $10~\%$--$15~\%$ (see also Fig.~\ref{fig_interf_voph}). For time-computational reasons, we only produced the synthetic visibilities for all the time-steps belonging to the best-fit photometric model. 

We note that the present work does not attempt to find the best pulsational single-cycle model fit to the set of three observational epochs (three different phases above mentioned), to derive a model of the full cycle; in fact we do not know if there is any model that would fit a full cycle of V~Oph in this model. We have, instead, used the model structures from the entire set of time-steps in the one model best fitting the photometric data to find the one atmospheric structure that best fits the observations at each specific epoch -- thusly making a semi-empirical determination of the parameters of the star's atmosphere at that epoch. But we do not intend, nor claim, to have found a model that corresponds to the star throughout its cycle. We have simply used this semi-empirical fitting to assess the differences in atmospheric structure from one phase to another in this cycle.

The interferometric $\chi^2$ values of the best-fit time-steps (Table~\ref{tab_param_dyn}) are provided for each of the three observed phases, to guide the discussion. For readability of the figures involving model visibilities, only the best time-step is shown; we show in Fig.~\ref{cycle_to_cycle} an example of figures of all the time-steps of this model.

In the following paragraphs, we present the results of the comparison of the dynamic atmosphere models with the photometric and interferometric archive data of V~Oph.

\section{Results}\label{results}


\subsection{Photometry}\label{photom}

The best-fit models for photometry of V~Oph resulted, at first, in a model without mass loss. Since this star displays loss of mass in the literature (see Table~\ref{table_starsparam}), we perform a selection a priori. We choose, among the grid of $540$ models, only the ones allowing for wind formation, that is, having a condensation factor $f_\text{c} > 0.2$. This results in a sub-grid of $168$ models, among which we performed our analysis. The best-fit models for photometry (``Photom'') and interferometry (``Interf'' at the three phases) are listed in Table~\ref{tab_param_dyn}, together with the initial best-fit windless model (``NO $\dot{M}$''). 

We would like to remark that this fit was done only with the aim of pre-selecting a model for the subsequent interferometric comparison. With this in mind, we noticed that, concerning photometry, the archival literature data longwards of $2~\mu$m contain only one observing date (single epoch). Data at shorter wavelengths often consist only of a small number of measurements, obtained during different light cycles. The only light curves available are the three ones in the $V$-band: \cite{asas, aavso, alf12} (see also Sect.~\ref{data} and Fig.~\ref{newlight}). Though, keeping our aim in mind, we noticed that the interferometric observations are not covered by the AAVSO and AFOEV light curves. Only in the case of the AAVSO light curve, photometric data are available at the three different phases of the interferometric observations (see also Figure~\ref{newlight}). The visual variability between the three epochs is: $V = 9.1$~mag at $\phi = 0.18$, $V = 10.4$~mag at $\phi = 0.49$, and $V = 9.3$~mag at $\phi = 0.65$, and we can consider it negligible with respect to the visual amplitude ($\Delta$V = $4.3$, see Table~\ref{table_starsparam}). We thus fit all the time-steps of the best-fit model to the average observed $V$-magnitude photometry. Section~\ref{discussion} reports an extensive discussion on the interferometric variability with the dynamic model atmosphere, and derivation of stellar parameters.

\begin{figure*}[!htbpbp]
\begin{center}
\resizebox{\hsize}{!}{
   \includegraphics[width=\hsize, bb=104 103 718 542, clip=true, angle=180]{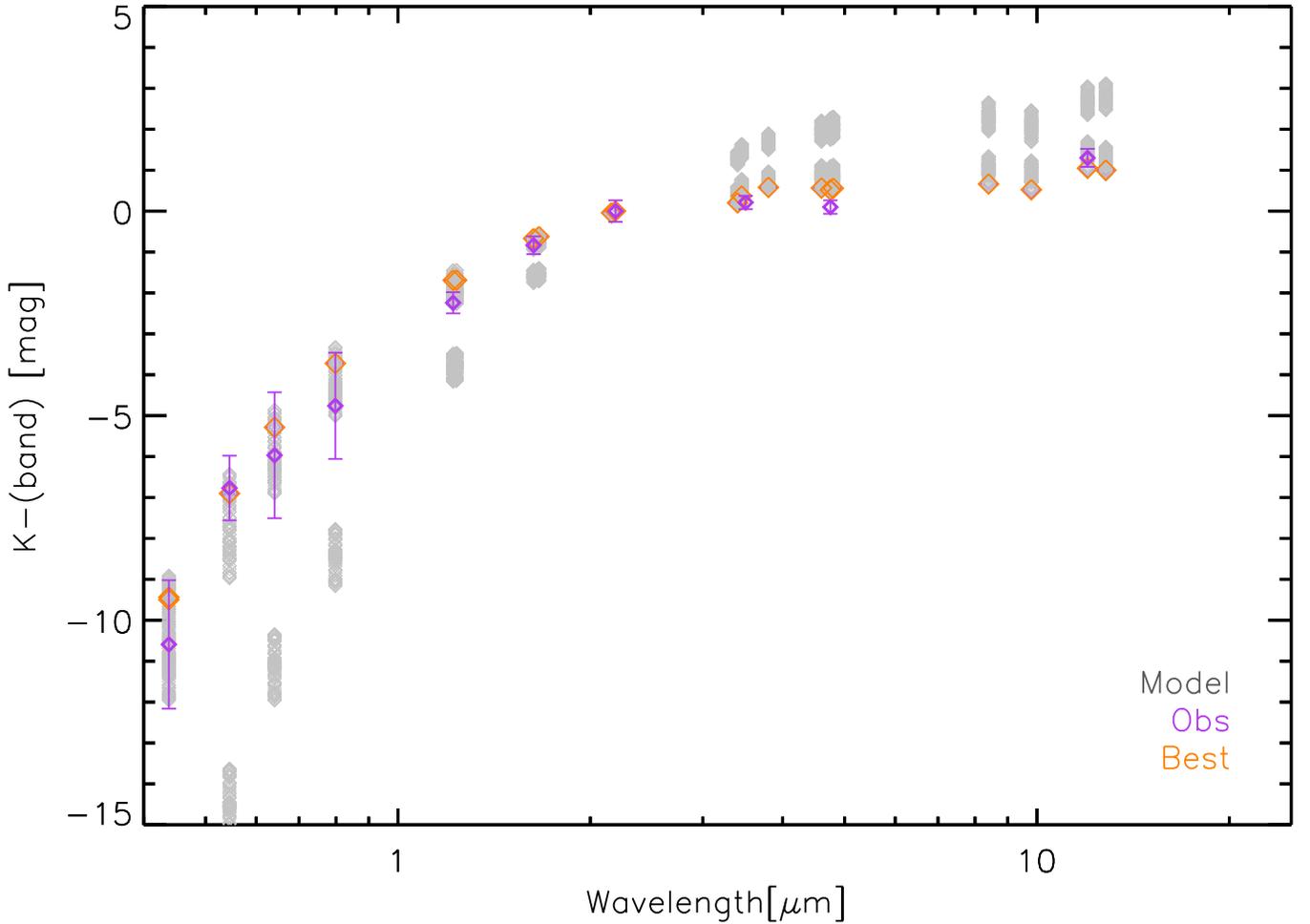}}
    \caption{\label{phot-voph} Photometric observations of V~Oph. Observations (purple diamonds) are compared to the synthetic photometry derived from the dynamic models (gray diamonds). All the phases of this best-fit model are plotted, and the best-fit time-step for photometry is marked with orange diamonds}.
\end{center}
\end{figure*}

The photometric data of V~Oph agree well with the model predictions (see Fig.~\ref{phot-voph}) at all wavelengths, within the error bars. The small differences at wavelengths shorter than $1~\mu$m, seen in \cite{rau17} for the two Mira stars of their sample (R~Lep and R~Vol), are not found in our fits of V~Oph. However, a literature spectrum covering this whole wavelength range is missing. In Sect.~\ref{disc_sedvis} we discuss the similarities and differences in terms of model and observed parameters. As previously noticed by \cite{rau15, rau17} in their spectral analysis of other sources, the photometric synthetic models tend to show a higher flux in emission at $\sim~14~\mu$m, which is not seen in the observed data. The origin of this feature predicted by the model, is due to C$_2$H$_2$ $\nu_5$ band, as well as HCN~$\nu_2$ (OH07), as mentioned also by \cite{loidl_phdthesis}. The good fit is confirmed by a $\chi^2$ of $1.72$ - see Table \ref{tab_param_dyn}.

\subsection{Interferometry}\label{interfer_results}

As mentioned already several time in this work, V~Oph shows interferometric variability. We thus fit its visibility profiles to the model independently, at each of the three phases.  

\begin{figure*}[!htbp]
   \centering   
   \includegraphics[width=0.8\hsize, bb=31 7 668 753, clip=true, angle=0]{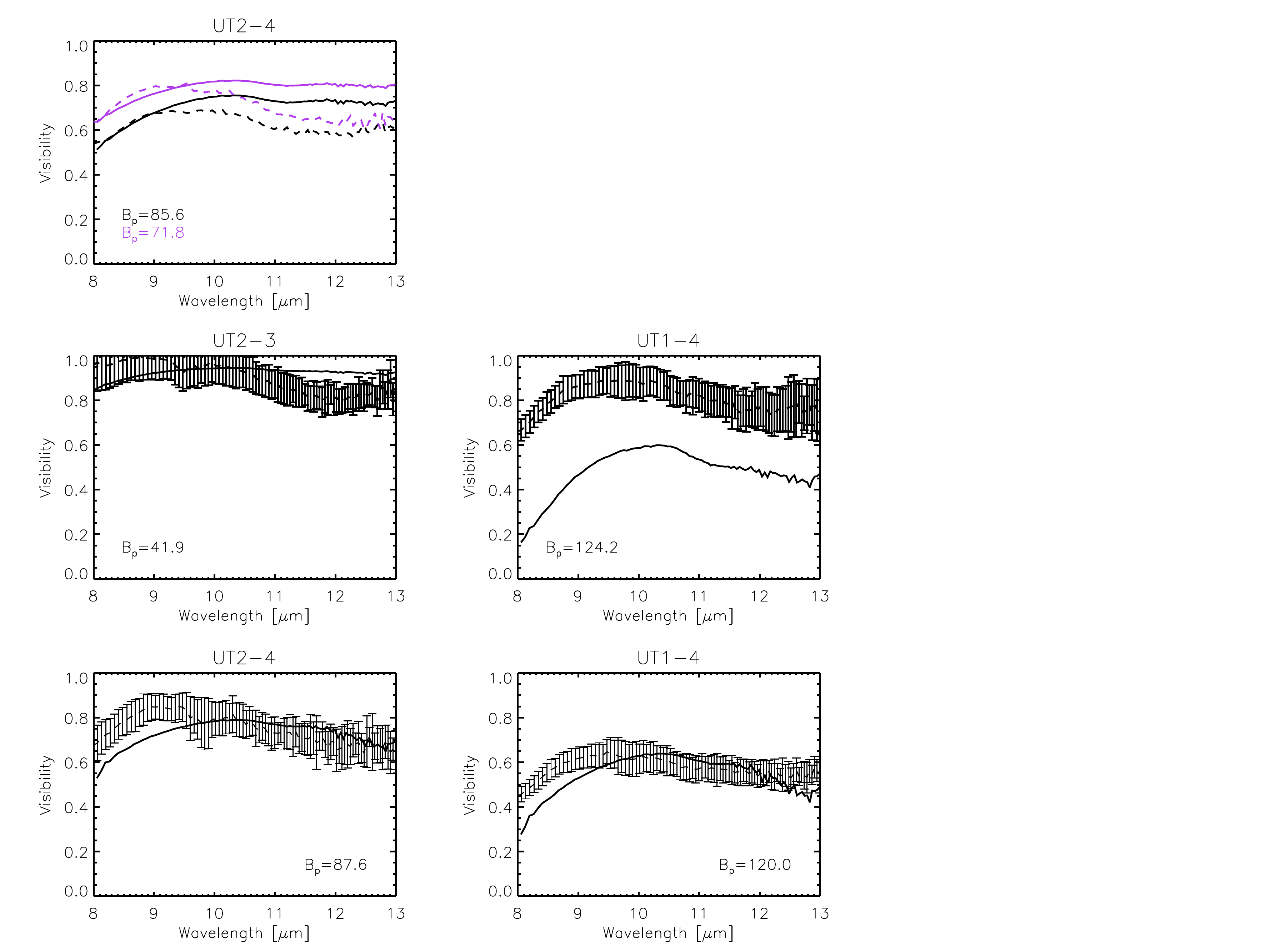}
    \caption{\label{fig_interf_voph} Visibility versus wavelength of V~Oph. The full lines represent the models, while observations are in dashed lines. The five panels show V~Oph visibility versus wavelength at the different baseline configurations of the Unit Telescopes (UT, indicated in the plot title), at the three phases: $\phi = 0.18$ (top panel), $\phi = 0.49$ (middle panels), $\phi = 0.65$ (bottom panels). The error bars are of the order of $10-15$~\%, not shown in the upper panel for a better readability of the figure.} 
\end{figure*}

\begin{figure*}[!htbp]
\begin{center}
\resizebox{\hsize}{!}{
\includegraphics[width=0.7\textwidth, bb=67 58 703 546, angle=180]{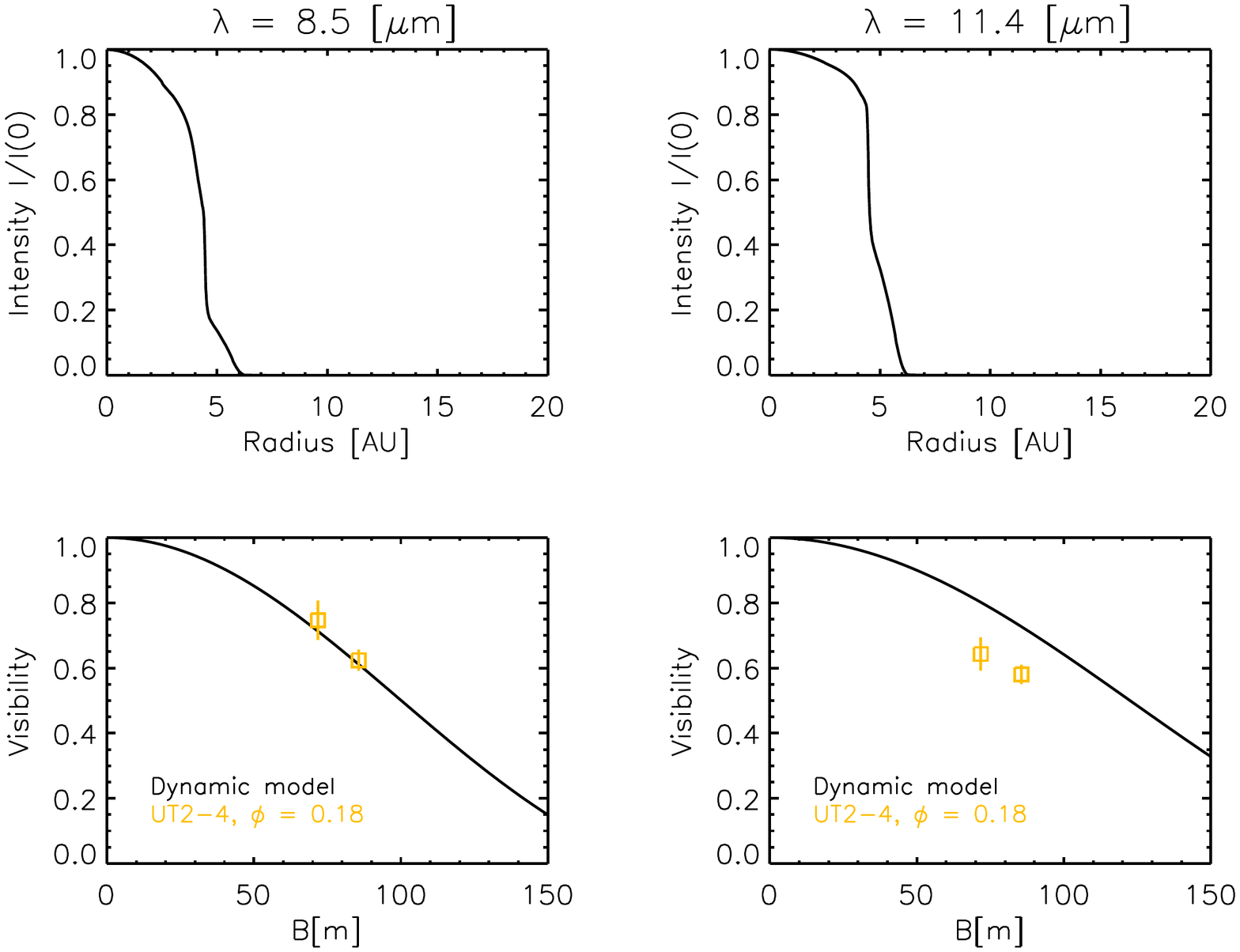}}
\caption{Interferometric observational MIDI data of V~Oph at phase $\phi = 0.18$, compared with the synthetic visibilities calculated from the DARWIN models. The two upper panels show the intensity profiles at two different wavelengths: $8.5~\mu$m and $11.4~\mu$m. The two lower panels illustrate the visibility versus baseline at these two wavelengths. The synthetic models are shown as black lines, and the colored symbols are the MIDI visibilities at different baselines configurations.}
\label{interf-voph-visbase1}
\end{center}
\end{figure*}

\begin{figure*}[!htbp]
\begin{center}
\resizebox{\hsize}{!}{
\includegraphics[width=0.7\textwidth, bb=67 58 703 546, angle=180]{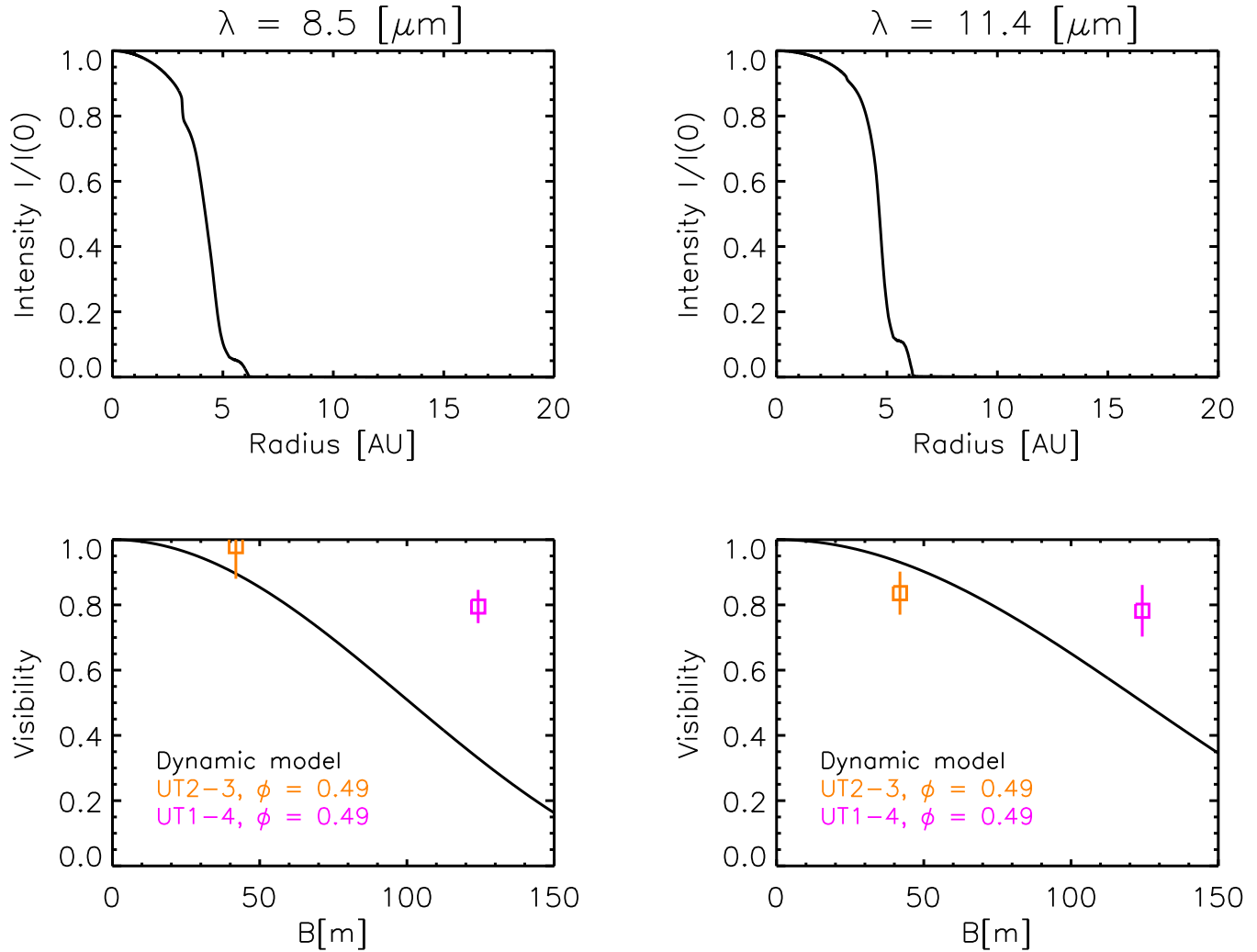}}
\caption{As Fig.~\ref{interf-voph-visbase1}, but at phase $\phi = 0.49$}.
\label{interf-voph-visbase2}
\end{center}
\end{figure*}

\begin{figure*}[!htbp]
\begin{center}
\resizebox{\hsize}{!}{
\includegraphics[width=0.7\textwidth, bb=67 58 703 546, angle=180]{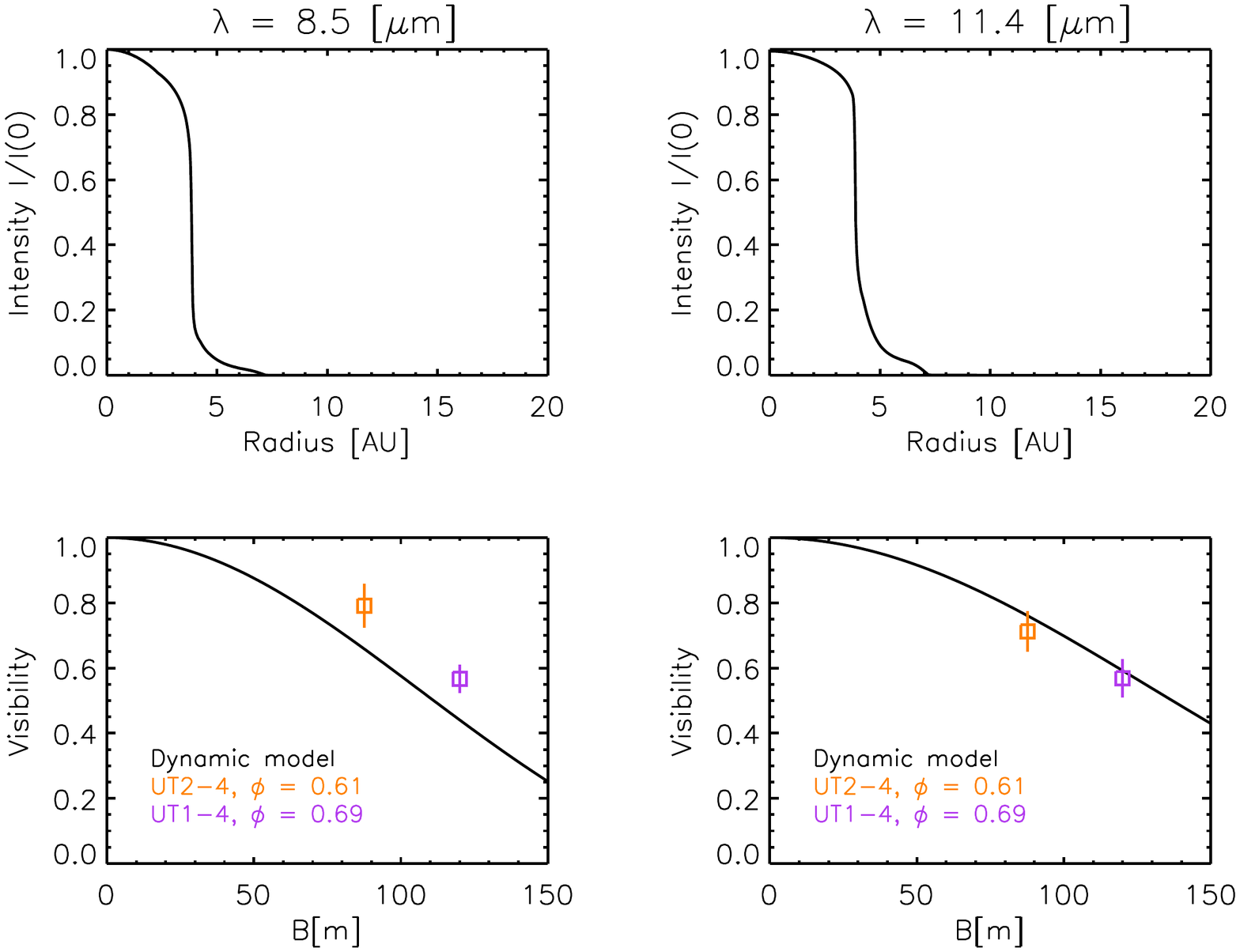}}
\caption{As Fig.~\ref{interf-voph-visbase1}, but at phase $\phi = 0.65$.}
\label{interf-voph-visbase3}
\end{center}
\end{figure*}

V~Oph interferometric data are shown, for each of the three phases, as visibility vs.~wavelength in Fig.~\ref{fig_interf_voph}, and as visibilities vs.~baselines in the lower panels of Figures~\ref{interf-voph-visbase1}, \ref{interf-voph-visbase2}, and \ref{interf-voph-visbase3}. The typical shape of carbon-rich AGB stars in the mid-infrared is noticeable in the visibility spectra, indeed the visibility versus wavelength go up from $\sim~8$ to $\sim~10.5~\mu$m. The emission from the C$_2$H$_2$~$\nu_4$ and $\nu_5$ bands makes the object appear larger than the star itself. Since the strength of the C$_2$H$_2$ bands becomes weaker from $8$ to $\sim10$~$\mu$m, the object's apparent size decreases from $8$ to $\sim10$~$\mu$m, resulting in the increase of the visibility. At longer wavelengths, the object's apparent size increases because of the presence of SiC dust around $11.3~\mu$m (see Fig.~\ref{fig_interf_voph}), resulting in the decrease in the visibility.

Table~\ref{tab_param_dyn} shows the fit at the three different phases. 
The quality of the fit, indicated by the values of the $\chi^2$, is fair. However, comparison of the visibility as a function of wavelength, also as a function of baseline, shows that the models do not reproduce well the observations at all the baselines/phases: we notice some differences between observations and models in terms of visibility shape (e.g.~at baselines $85.6$~m, $71.8$~m, $41.9$~m, see Fig.~\ref{fig_interf_voph}, upper panel and middle-left panel) and absolute visibility level (e.g.~at baseline $B_\text{p} =124.2~m, \phi=0.49$, see Fig.~\ref{fig_interf_voph}, middle-right panel); these differences are discussed in details in Sect.~\ref{discussion}.

\section{Discussion}\label{discussion}

\subsection{The SED, and the interferometric visibilities}\label{disc_sedvis}

The dynamic atmosphere models reproduce well the literature photometric data, and quite well the archive MIDI interferometric ones, even though the latter with discrepancies at the longest baseline. The best-fit model for V~Oph is characterized by a relatively compact atmosphere, with gas and dust shells less pronounced than the previous models for carbon-rich Miras (e.g.~\citealp{rau15}, \citealp{rau16}, \citealp{rau17}). The best-fit model for V~Oph shows episodic mass loss, and its mass-loss rate is $\sim~20$~times higher than the observed values reported in literature (see Table~\ref{table_starsparam}). It should be noted that the model mass-loss rates are computed as averages over pulsation periods as described in Sect.~$2.3$ of \cite{erik14}. Similar higher values for the mass-loss rates were found in \cite{rau17} with the best-fit model mass-loss rate being $\sim10$ times higher than the literature value for the semi-regular star \object{Y~Pav}, and $\sim5$ times for the irregular star \object{X~TrA}.

In \cite{rau17} we noticed that the models best-fit the Mira stars \object{R~Lep} and \object{R~Vol} have a more pronounced shell-like structure (i.e.~the discontinuous, step-like structure of the intensity vs.~radius plot) than for the non-Mira stars, and their visibility vs.~baselines profiles decline, leveling off at longer baselines (see Fig.~$6$ in \citealp{rau17}). The best-fit models for V~Oph are characterized by compact intensity profiles, as Figures~\ref{interf-voph-visbase1}, \ref{interf-voph-visbase2}, \ref{interf-voph-visbase3} show. These differences could be related to the atmosphere of V~Oph, which is probably less extended, and to its atmosphere, which shows signs of weak shell-like stratification. This suggests that the atmosphere of V~Oph, albeit its classification as a Mira star, might be similar to those of non-Mira stars (this is discussed in Sect.~\ref{voph_semireg_vs_mira}).


The best-fit model of V~Oph interferometric data shows, as in the case of all the non-Mira stars in \cite{rau17} (see their Figure~$4$), high visibility levels but no slope, i.e.~the model visibilities do not increase with wavelength.  We could explain this different behavior by examining the density structure (in terms of density vs.~radius, discussed in Appendix \ref{appendix_windless}, Fig.~\ref{density_overplot}) of the best-fit time-steps at each phase. We notice indeed that the the density profile in V Oph is smoother (i.e., the density falls off slowly with radius; see Figure~\ref{density_overplot}) compared to the case of the Mira stars previously studied (i.e.~\object{RU~Vir}, \object{R~Lep}, \object{R~Vol}). As hypothesized by \cite{rau17}, such smoother density distribution does not produce the slope difference seen in the Mira models of R~Lep and R~Vol. 
From this result we could portray the atmosphere of V~Oph to be characterized by a less pronounced shocks, which might result from a different cooling function or different dust formation parameters. 

One possible way of interpreting these results is that the shock waves producing dust change at different cycles. A different interpretation could be related to the choice of selecting \textit{a priori} only those models producing mass loss (see Sect.~\ref{photom}). To verify our choice, we check whether or not the model without mass loss (which was originally best-fitting the photometric models, see Sect.~\ref{photom}) reproduces the visibilities well (this was the case for e.g.~\object{R~Scl} in \cite{wittkowski17}). We show the results of this experiment in Appendix~\ref{appendix_windless}, which shows that the \textit{windless} model cannot reproduce the data as well (see Figure~\ref{visib_windless}).

However, an alternative explanation could be that such smooth, and not extended, intensity profiles could be caused by, instead of large-amplitude stellar pulsation alone, some other mechanism operating in the lower part of the atmosphere, such as Alfv\'{e}n waves propagating at different speeds \citep{vidotto06, suzuki06}. This possibility is discussed in Sect.~\ref{discuss_alfven}.  Another possible scenario for the behavior of V~Oph atmosphere is that the star could be more similar to semi-regular/irregular stars than to Mira stars. This hypothesis is discussed in Sect.~\ref{voph_semireg_vs_mira}.

We discuss below, separately, the shape and the level of the visibility profiles:

\begin{itemize}
\item \textbf{Level of visibility profiles} Overall, the visibilities fit show that at phase $0.18$ (Fig.~\ref{fig_interf_voph}, upper panel) the model can reproduce well the observations from $8$ until $11.2~\mu$m for baseline $71.8$~m, and until $9.4~\mu$m for baseline $85.6$~m, within the error bars (which are not shown in the corresponding panel for clarity reasons). At phase $0.49$ (Fig.~\ref{fig_interf_voph}, middle panel) the model at baseline $41.9$~m (left) are within the observations error bar until $\sim11.1~\mu$m, while for baseline $124.2$~m (right) the models do not reproduce the observations in the entire wavelength range. Phase $0.65$ (Fig.~\ref{fig_interf_voph}, lower panel) shows a good fit of visibilities vs.~wavelength from $9.3$ until $12.5~\mu$m for baseline $87.6$~m (left) and from $10$ until $13.0~\mu$m for baseline $120.0$~m (right).

\item \textbf{Shape of visibility profiles} This shape of the visibility profiles (i.e., the visibility at a given wavelength as a function of spatial frequency or baseline), is not always reproduced by the models: we observe that the synthetic visibility profile remains flat after $10.5~\mu$m for $\phi = 0.18$ and $\phi = 0.49$ $B_\text{p} =41.9$ (upper and middle-left panels in Fig.~\ref{fig_interf_voph}). In the latter, we notice that at the long baseline ($B_\text{p}  = 124.2$~m) the shape of the visibilities is reproduced well by the models, but not the level (as encountered in \citealp{rau17}), and this can be seen also in the visibilities vs.~baselines at the same phase (Fig.~\ref{interf-voph-visbase2} right panel). We notice that the long baseline of $\phi = 0.65$ ($B_\text{p}  = 120.0$~m, lower right panel in Fig.~\ref{fig_interf_voph}) the visibility profile fits better than at the long baseline of $\phi = 0.49$, both in terms of visibility vs.~wavelength, and especially of visibility vs.~baselines (Fig.~\ref{interf-voph-visbase3}, right panel).

\end{itemize}

Concerning the visibilities vs.~baselines, we notice the same trend as in \cite{rau17}: in general the fit is better at $8.5~\mu$m than at $11.4~\mu$m, exception made for $\phi = 0.65$ (Fig.~\ref{interf-voph-visbase3}). The latter could be due to the fact that, for such compact (less extended) intensity profiles, the effect of the SiC inclusion in the models could be more pronounced than for the extended profiles (see \citealp{rau17}, their Figs.~$6$ vs.~Fig.~$7$, and discussion therein).

An explanation of the discrepancies found in fitting the longest baseline $B_\text{p} = 124.2$~m at phase $0.49$ could be related to e.g.~regular or irregular dust obscuration events (see e.g., \citealp{clayton12} for R Crb stars), or by deviation from spherical symmetry due to a possible binary star which create an elongated dust envelope or disk. However, the $B_\text{p}$ and PA of phase $0.49$ ($B_\text{p} =124.2$~m, PA=$65.6$) and phase $0.65$ ($B_\text{p} =120.0$~m, PA=$66.0$) are very similar. Hence, while non-spherical nature of the object cannot be excluded, it is unlikely to explain the bad fit of phase $0.49$, $B_\text{p} =124.2$~m. We also checked if there is significant cycle-to-cycle variation at phase $0.49$ in the model visibility, which might explain the poor fit, and for this we refer to Section~\ref{appendix_binary}, where we report a substantial cycle-to-cycle variability among all the time-steps of V~Oph at the baseline $B_\text{p} =124.2$~m.


\subsection{The cycle-to-cycle variations at phase $0.49$}\label{appendix_binary}

\begin{figure*}[!htbp]
\begin{center}
\includegraphics[angle=180,  width=0.45\hsize]{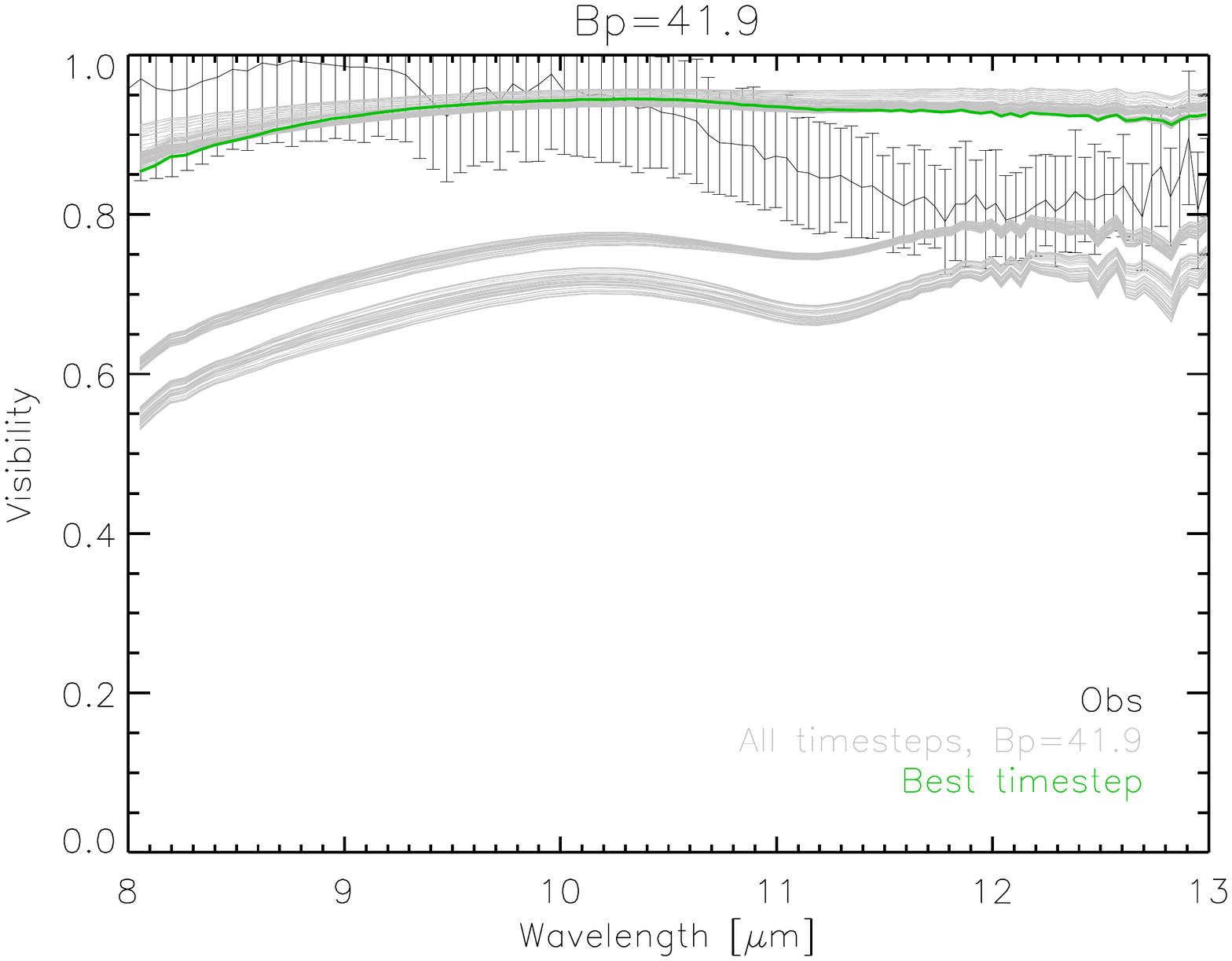}
\includegraphics[angle=180,  width=0.45\hsize]{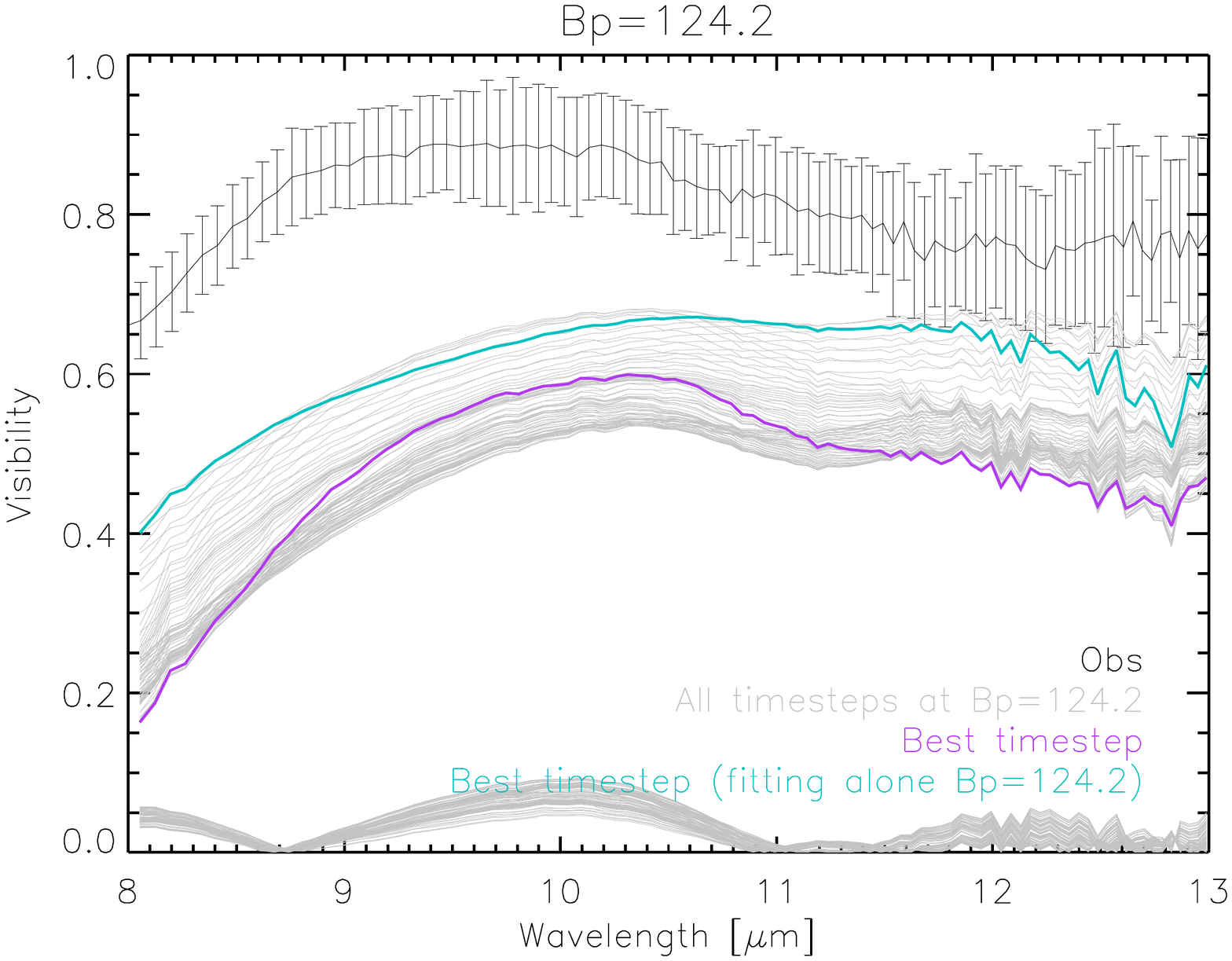}
\includegraphics[angle=180,  width=0.45\hsize]{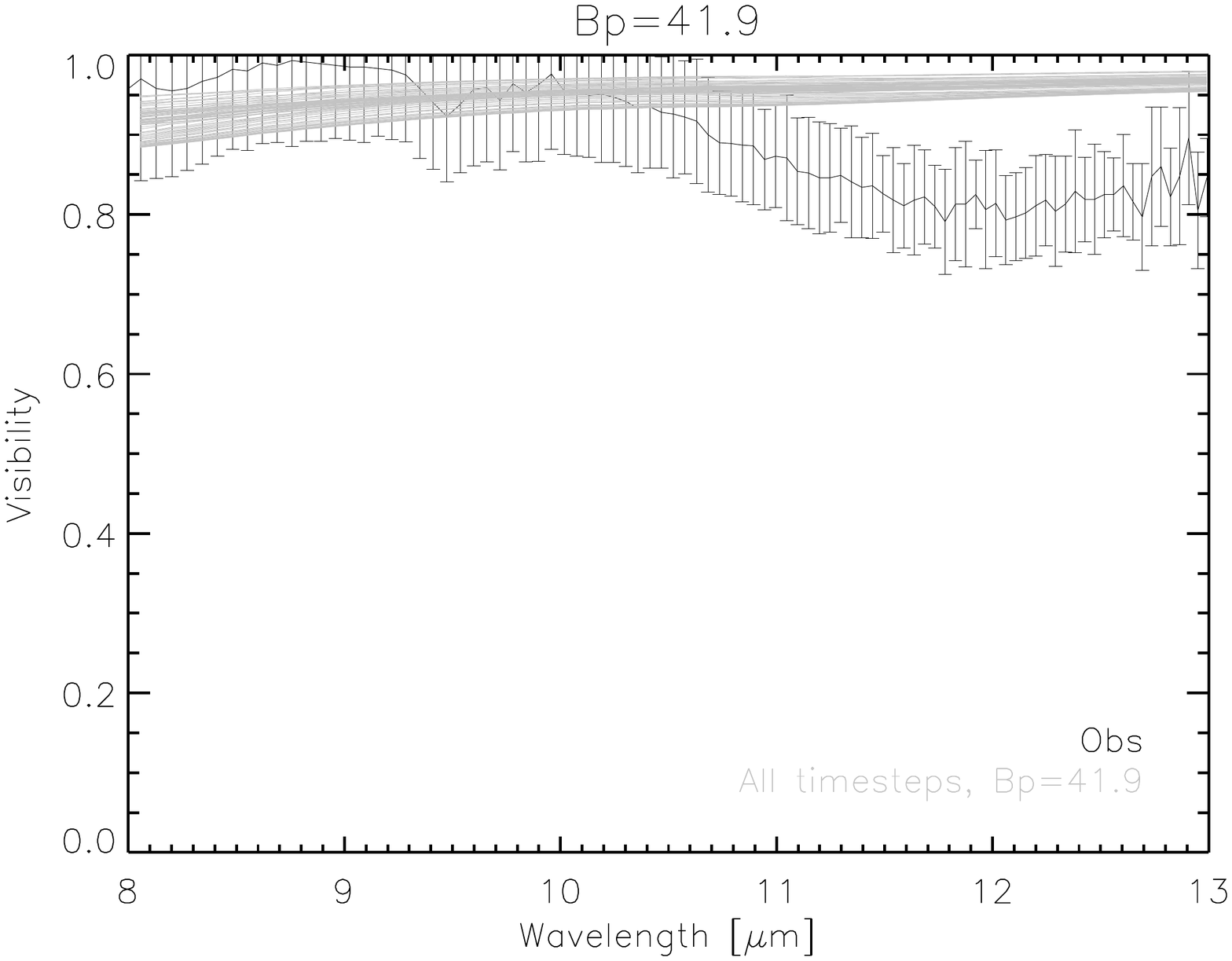}
\includegraphics[angle=180,  width=0.45\hsize]{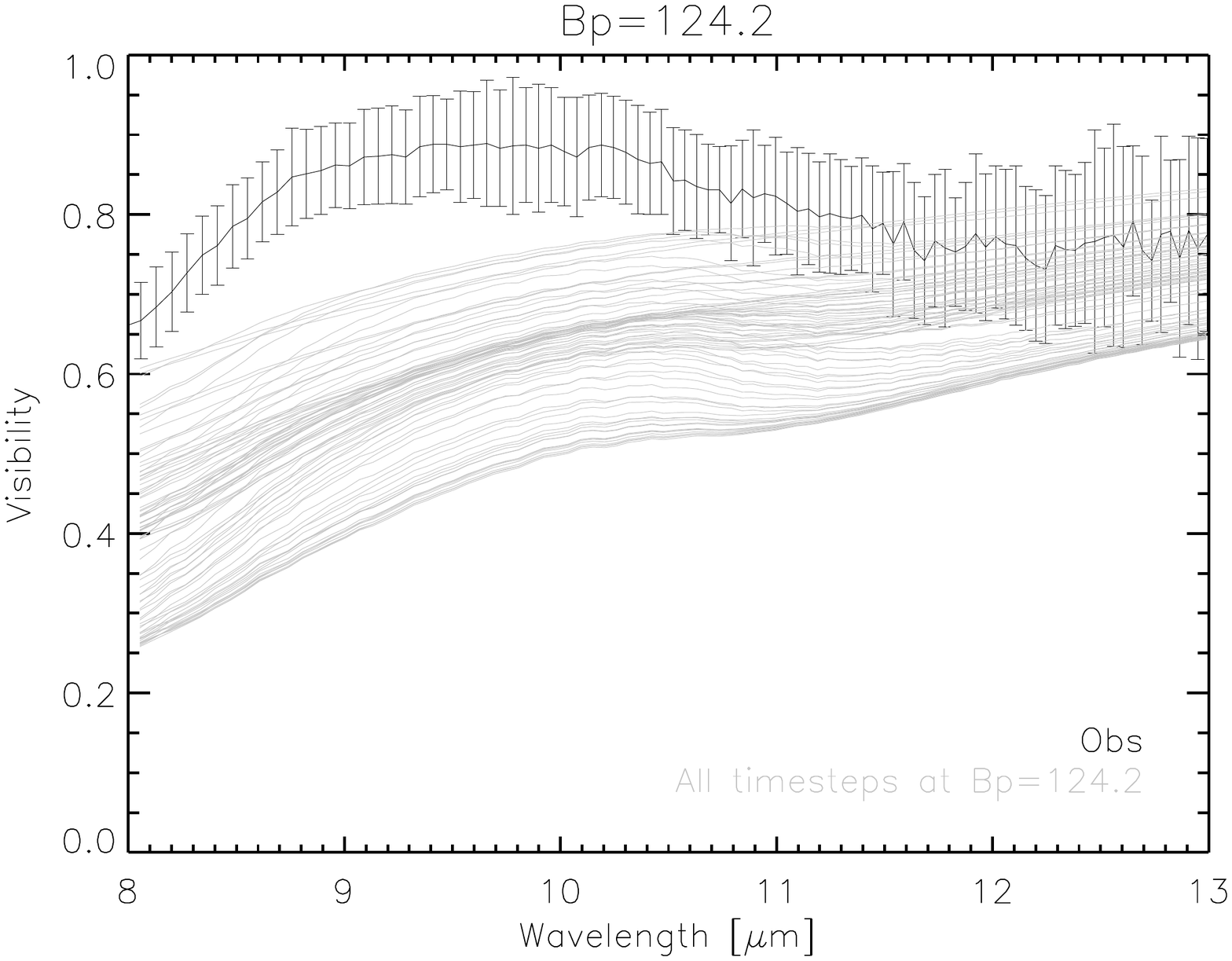}
\caption{\textbf{Top}: Cycle to cycle variation of the models with wind, at phase $\phi = 0.49$, for the baseline $B_\text{p} = 41.9$~m (left panel) and $B_\text{p} = 124.2$~m (right panel). The gray lines are the visibility versus wavelength at all the time-steps of V~Oph best-fit model with wind. The observation with relative errors are in black. And the best-fit time-step is in green for $B_\text{p} =41$~m, and purple for $B_\text{p}=124$~m. The cyan line at $B_\text{p} = 124.2$~m is the best-fit time-step resulting from fitting this baseline alone to the set of time-steps. \textbf{Bottom}: Same as the top panels, but for the \it{windless} model.}
\label{cycle_to_cycle}
\end{center}
\end{figure*}

A significant cycle-to-cycle variation at phase $0.49$ could impact the fit of the visibility spectra. We verify such scenario showing the visibility vs.~wavelength at phase $0.49$ for all the time-steps of V~Oph best-fit model with wind (projected baselines: $B_\text{p}=41.9$~m and $B_\text{p}=124.2$~m, see Fig.~\ref{cycle_to_cycle} upper left and right panels respectively). We report indeed a substantial cycle-to-cycle variability among all the time-steps of V~Oph at the baseline $B_\text{p}=124.2$~m. 

We would like to underline that we compare the model prediction to both baselines data at the same time (but not averaging them). The resulting best-fit time-step, reported in Fig.~\ref{cycle_to_cycle} (see also Table~\ref{tab_param_dyn}), is marked in green line for $B_\text{p}=41.9$~m and purple line for $B_\text{p}=124.2$~m. The cyan line represents the best-fit time-step obtained fitting \textit{only}~$B_\text{p} =124.2$~m to the grid of time-steps (vs.~fitting both baselines at the same time), and it corresponds to a $\chi^2$ of $9.95$. Instead, fitting only $B_\text{p} =41.9$~m alone to the grid of time-steps of the best-fit photometric model would give a $\chi^2$ of $1.51$.

For comparison reasons, and for completeness, we also show all the time-steps of the \textit{windless} model at the two baselines (see Fig.~\ref{cycle_to_cycle}, lower panels). We notice that at $B_\text{p} =124$~m the visibility level increases of $\sim0.2$ at the shorter wavelength, but the shape is not well reproduced, concluding that comparing the data to the windless model does not improve the quality of the fit.

\subsection{Fundamental stellar parameters compared to the literature, and evolutionary tracks}\label{evol_tracks}
        
From the best-fit models at the three phases we derive a number of parameters as listed in Tab.~\ref{tab_param_dyn}. These values refer to the hydrostatic initial structure of the model (see also \citealp{nowotny05}). We then calculate the Rosseland diameter ($\theta_\text{Ross}$), to be able to compare the fundamental parameters of V~Oph to the ones found in literature; this is the diameter corresponding to the distance from the center of the star to the layer at which the Rosseland optical depth equals $2/3$. We calculated also the corresponding effective temperature $T_\text{Ross}$, i.e.~the temperature of the time-step at this Rosseland radius. Following the Stefan-Boltzmann law we can compute the Rosseland luminosity $L_\text{Ross}$. Furthermore, from the literature photometric data, we derive the bolometric luminosity $L_\text{bol}$, a diameter $\theta_{\text{(V-K)}}$, using the diameter vs.~$\text{(V-K)}$ relation of \cite{vanbelle13}, and its corresponding effective temperature \textit{T}($\theta_{\text{(V-K)}}$). The error on the luminosity is assumed to be approximately $40$~\%, on the basis of the distance uncertainty. (The uncertainties are known to be underestimated by up to $\sim30$~\% for bright stars \citealp{gaiaparallaxes}). The errors of the temperature are estimated through the standard propagation of error. We list all these parameters in Table~\ref{tab_theta}, for the three different phases. No K-band diameter is available for this target. We note that the observed and model gas velocities (see respectively Table~\ref{table_starsparam} and Table~\ref{tab_param_dyn}) agree quite well.

\begin{figure*}[!htbp]
\centering
\includegraphics[width=0.7\hsize, bb=91 103 702 539, angle=180]{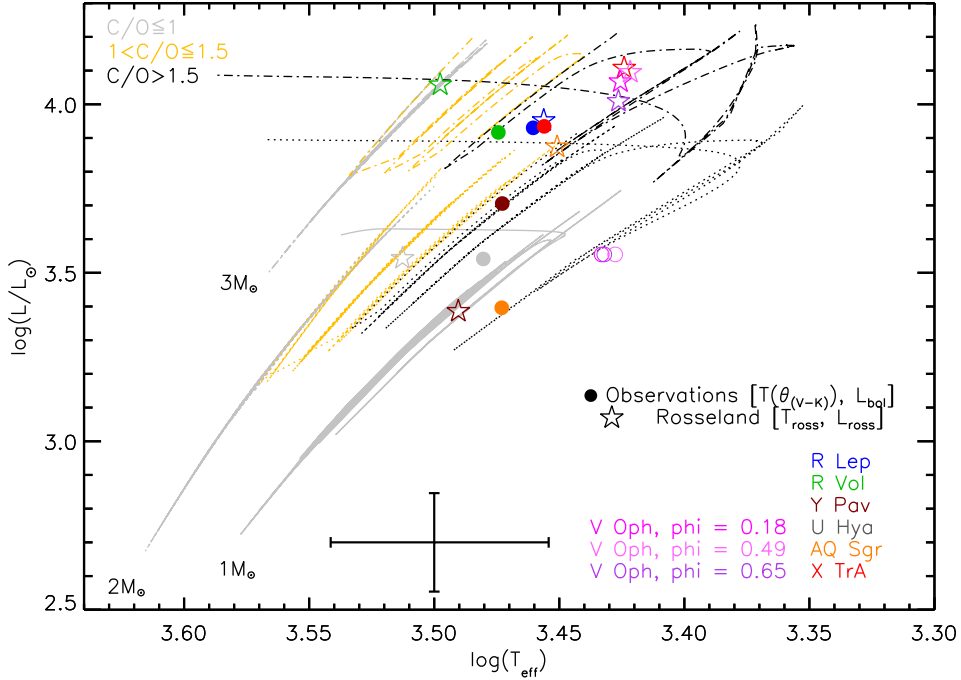}
\caption{\label{evolut-track} Hertzsprung-Russel diagram zooming on the Asymptotic Giant Branch part. The lines show solar metallicity evolutionary tracks from \cite{marigo13}. The different lines colors display the regions of: Oxygen-rich stars in gray ($C/O<1.0$); C-rich stars in yellow ($1.0<C/O\leqslant~1.5$); stars with $C/O~>~1.5$ are in black. The numbers indicate the mass values at the beginning of the thermal pulsing (TP)-AGB. The track with $2~$M$_{\odot}$ is plotted with a dotted line, and the one with $3~$M$_{\odot}$ in a dot-dashed line. Different symbols and colors refer to the luminosity and effective temperature, estimated through the comparison in this work of the models with photometric and interferometric observations. A typical error-size bar for the observation estimates is shown in the lower side of the figure. We note that the circles representing V~Oph location on the tracks are open, unlike the ones for the other stars that are full, to make them better distinguishable, otherwise they would overlap.}
\end{figure*}

We compare the evolutionary stage of V~Oph with ones for previously studied Mira and non-Mira stars (see Fig.~\ref{evolut-track}). We place temperatures and luminosities in thermally-pulsing (TP) AGB evolutionary tracks from \cite{marigo13}, illustrating the Thermal Pulse (TP) AGB tracks for three choices of the initial mass on the TP-AGB: $1.0, 2.0$~and~$3.0~$M$_{\odot}$. The first TP is extracted from the PARSEC database of stellar tracks \citep{bressan12}, and, starting from this, the TP-AGB phase is computed, until the whole envelope is removed by stellar winds. The TP-AGB sequences are selected with an initial scaled-solar chemical composition: the mass fraction of helium Y is $0.273$, and the one of metals Z is $0.014$. To guarantee the full consistency of the envelope structure with the surface chemical abundances, which may significantly vary due to the $3^{\text{rd}}$ dredge-up episodes and hot-bottom burning, the TP-AGB tracks are based on numerical integrations of complete envelope models in which, for the first time, molecular chemistry and gas opacities are computed on-the-fly with the {\AE}SOPUS code \citep{marigoaringer09}. For further details on the calculations of the evolutionary tracks we refer to \cite{bressan12, marigo13, marigoaringer09}.

We note that the TP-AGB model for $M=1.0$~M$_{\odot}$ does not experience the third dredge-up, hence remains with $C/O~<~1$ until the end of its evolution. Conversely, the model with $M=2$ and $M=3$~M$_{\odot}$ experiences a few third dredge-up episodes that lead to reach $C/O~>~1$, thus causing the transition to the C-star domain. The location of the observed C-stars in the H-R diagram, as well as their $C/O$ ratios, appear to be nicely consistent with the part of the TP-AGB track that corresponds to the C-rich evolution. It is worth noting that the current mass along the TP-AGB track is reduced during the last thermal pulses, which supports (within the uncertainties) the relatively low values of the mass ($\sim 0.75-1.0~$M$_{\odot}$) assigned to some of the stars studied by \cite{rau17} and reported as well for comparison in Fig.~\ref{evolut-track} through the best fitting search on the DARWIN models dataset.

The model mass resulting from the dynamic atmosphere model is $1.5~$M$_{\odot}$ (see Table~$2$, second column). Such low current mass has been reduced during the last termal pulses along the TP-AGB tracks; hence, within the uncertainties, this value agrees with the one derived from the observations (open circles), which place the star on a $2.0~$M$_{\odot}$ track before a TP phase, while the Rosseland calculations place V~Oph in the vicinity of a $3.0~$M$_{\text{sun}}$ track. We also notice a good agreement with the mass derived using the \cite{wood15} W$_{JK}$ index and period (see Sect.~\ref{voph_semireg_vs_mira}).


We notice that the temperature derived from the $\text{(V-K)}$ relation and the bolometric luminosity of V~Oph resemble better the ones of semi-regular stars, than the ones of the other Mira stars. 

The differences between the luminosity and estimations, $L_\text{Ross}$ and $L_\text{bol}$, are not within the error bars. This may be related to the above mentioned episodic mass-loss of the best-fit model. Literature values of luminosity can be found in \cite{mcdonald17}, and they agree within the uncertainties, to our estimate, considering the differences in the data sets and methods (Gaia DR2 is not refined yet for the most luminous stars\footnote{E.g., \url{https://gea.esac.esa.int/archive/documentation/GDR2/Data_processing/chap_cu6spe/sec_cu6spe_qa/ssec_cu6spe_val.html}, \cite{katz19}, and \cite{BailerJones18}}). We underline the very good agreement between $T_\text{Ross}$ and the purely empirically determined \textit{T}($\theta_\text{(V-K)}$), as in the case of \cite{rau17}.

We observe that the diameters $\theta_\text{Ross}$ and $\theta_{\text{(V-K)}}$ in Table~\ref{tab_theta} agree extremely well, despite the episodic mass loss of the best-fit model.

\begin{table*}
\centering
\small
\caption{\label{tab_theta} Observed and calculated temperatures and diameters.}
\begin{tabular}{llllllll}
\hline
\hline 
Target  & ${\phi}$ & $\theta_{\text{(V-K)}}$~\textsuperscript{a} & $T_{\theta_\text{(V-K)}}$ & $\theta_\text{{Ross}}$~\textsuperscript{b}  & R$_\text{{Ross}}$  &   $L_\text{{Ross}}$ &  $T_\text{{Ross}}$  \\
     &    &      [mas]               &   [K]         &   [mas]  &   [R$_{\odot}$]       & [L$_{\odot}$]  &  [K] \\
\hline
V~Oph   & 0.18 &         $3.45$          & $2710 \pm 332$        & 3.23     &  253     &         $11669$  &    $2664$\\
V~Oph   & 0.49 &         $3.54$          & $2677 \pm 328$        & 3.39     &  266    &         $12442$  &    $2640$ \\
V~Oph   & 0.65 &         $3.46$          & $2704 \pm 331$        & 3.01     &  236      &         $10204$  &   $2669$\\
\hline
\hline
\end{tabular}\\
\textbf{Notes}. (a) Relation from \cite{vanbelle13}. (b) $\theta_\text{Ross}$ is the Rosseland diameter of the best-fit time-step of the corresponding best-fit model. 
\end{table*}

\subsection{A comparison with \cite{ohnaka07}}\label{discuss_comparison}
In this section we compare the results obtained by OH07 with their modeling, with the ones in the present work derived from the modeling with self-consistent dynamic atmosphere models. OH07 modeling consisted of a so called ``MOLSPHERE'' model, with two separate molecular layers, and one layer of dust. The details of this models can be found in \cite{ohnaka07}.

OH07, with their simple layers modeling, estimate the dust layer to be at $\sim~2.5$~R$_{\ast}$, and calculate the stellar photospheric angular sizes (see Table~$3$ in OH07) at the three different phases. We converted these values in stellar radii. Those are reported in Table~\ref{table_radii}, together with the model photospheric radius (model size), and the dust radii, i.e.~the radii in the models at which the dust starts to condense, derived at each phase from our modeling.

The modeling of this paper improves the 
the estimation of the photospheric and dust radii of V Oph. This motivates future snapshot and imaging observations of this star in the mid- and near-IR, to observe if molecule and dust may co-exist in the atmosphere of this star. Studying the variation in temperature/radius of molecules and dust will help putting constraints on their distribution in the circumstellar environment of C-rich AGB stars (see also Section~\ref{conclu}).

\begin{table*}
\caption{\label{table_radii} Photospheric radius (marked with an asterisk) and dust layers extension, from the modeling of this paper. For comparison, values from OH07 are also reported.}
\centering
\begin{tabular}{lllll}
\hline\hline             
$\phi$ &  R$_{\ast}^{\text{OH07}}$&   R$_{\ast}^{\text{this paper}}$  &  R$_{\text{dust}}^{\text{OH07}}$ &   R$_{\text{dust}}^{\text{this paper}}$  \\
 &  [R$_{\odot}$]  &  [R$_{\odot}$]  & [R$_{\odot}$] & [R$_{\odot}$]    \\
\hline
0.18 & 352 &   479   &  880 &  780        \\
0.49 & 402 &   494   & 1005 &  853      \\
0.65 & 360 &   448   &   901 &  787   \\
\hline
\end{tabular}
\end{table*}

From Table~\ref{table_radii} we note that the value of the photospheric radii estimated in this work are, at each phase, $\sim~25~\%$ higher with respect to OH07 results. 
The difference with respect to OH07 could be due to the way in which the DARWIN models consider, in a self consistent way, the production of dust, i.e.~including radiation pressure on gas, which is not present in the simple layer model of OH07. The larger difference w.r.t.~OH07 in the stellar radii may result from the fact that stellar radius was not directly measured in the MIDI data.

\subsection{Discussion on the possible causes of the temporal variations of the outer atmosphere}\label{discuss_alfven} 

In this section we discuss three possible causes of the interferometric variability detected by OH07. As stated above, variability was observed by OH07 between the interferometric data at phase $\phi = 0.49$ and $\phi = 0.65$ (see Sect.~\ref{data}, and Table~$1$ in OH07). This variability means that OH07 observations of visibility vs.~wavelength, taken approximately at the same projected baseline length and position angles but $60$~days apart, indicate a temporal variation at certain wavelengths (see their Table~$1$, datasets \#3 and \#6). These observations clearly indicate a temporal variation at $8.3~\mu$m, $10~\mu$m, and $12.5~\mu$m.


\subsubsection{Alfv\'{e}n waves}
We investigate below if the prominent role in causing C$_2$H$_2$ molecular extended layers is due to the dust-driven winds triggered by large-amplitude stellar pulsation alone, or if other physical mechanisms could be called into question. 
The acceleration of this shell indeed can be driven by the stellar radiation pressure on molecules and dust, and/or pulsation and convection, and/or thermal or magnetic pressure. Recently \cite{hoefnerandfreytag19} show, modeling an M-type star with 3D CO5BOLD radiation-hydrodynamics simulations, that pulsation and convection are likely mechanisms for producing observed clumpy dust clouds. However, here we provide a hypothetical, alternative scenario.

Using the results from OH07, the velocity of motion of a molecular shell of C$_2$H$_2$ from $\sim1.4$~R$_{\ast}$ to $\sim1.7$~R$_{\ast}$ at the time scale of $60$~days represented by the time difference between the phases $0.49$ and $0.65$, is $\sim20$~km/s. From the model of wind acceleration via radiative pressure on dust best fitting our observations, the gas velocity is $v_\text{e}\sim8$~km/s (see Table~\ref{tab_param_dyn}), i.e.~$2.5$~times less than observed in C$_2$H$_2$ shells. Gas motions with velocities of this order have been seen in the atmospheres of many AGB stars (e.g., \citealp{hinkle78}, \citealp{vlemmings17}; \citealp{vlemmings18b}), and these are interpreted as an effect of stellar pulsation. This is certainly a plausible interpretation. However, there are not sufficient observations in terms of time coverage of the whole stellar cycle, to confirm that the $20$~km/s flow is solely due to pulsation. In addition, e.g.~\cite{vlemmings18}, referring to \cite{falcetagoncalves02}, as well as several other authors (see also Sect.~\ref{intro} and the end of this subsection), mentions that magnetic field could help levitate material off the stellar surface via Alfv\'{e}n waves.

The gas in the shell is a weakly ionized gas that can support the propagation of Alfv\'{e}n waves via ion-neutral collisions. The propagation of such waves in weakly ionized star-forming environments and molecular clouds is invoked to explain gas~\&~dust acceleration as they are less subject to damping as compared to acoustic and compressible magnetohydrodynamic (MHD) waves (e.g.~\citealp{jatencopereira14}, \citealp{mckee95}, \citealp{ballester18}). This suggests that the additional momentum could be exerted by some mechanical flux. The surface magnetic field has been measured on a number of giant and supergiant stars and post AGB stars at the strength of a few Gauss (e.g.~\citealp{sabin15, vlemmings14, duthu17, vlemmings18}); this suggests that rigorous convection could excite Alfv\'{e}n waves as on the Sun. 


V~Oph molecular shell contains gas consisting of mainly hydrogen, and C$_2$H$_2$. From \cite{cherchneff12} -- see their Table $2$ -- the total gas density for a carbon star at $1.5$~R$_{\ast}$ is n$_{gas} = 8.24\cdot10^{12}$~cm$^{-3}$. The mass density of this gas is about: $\rho=n_\text{gas}\cdot M_\text{gas} \approx 6\cdot10^{-11}$~g~cm$^{-3}$. However, this gas is partially ionized, and we can consider the ionization fraction for a carbon star of the order of $7.8\cdot10^{-5}$ \citep{groenewegen97}. With the gas mass density $\rho$ above mentioned, and this ionization fraction, we obtain a mass density of the ionized gas: $\rho_{\text{ionized}}=4.68\cdot10^{-15}$. Ionization fraction limits are given also by \cite{spergel83}: $<4\cdot10^{-5}$; and \cite{drake91}: $<10^{-4}$. These correspond to, respectively, $\rho_{\text{ionized}} < 2.4\cdot10^{-15}$, and $\rho_{\text{ionized}} < 6\cdot10^{-15}$.




Maser polarization observations in AGB stars provide constraints on the magnetic field (see \cite{vlemmings18}, Fig.~$1$). Assuming that the wind from V~Oph is formed from an open radially diverging magnetic field (see e.g.~\citealp{airapetian00, airapetian10}), then from the conservation of the magnetic flux, $B\propto~r^-2$ (also consistent with Fig.~$1$ of \citealp{vlemmings18}). Then the magnetic field strength at the distance of the molecular layer should be $\sim10$~G. This is consistent with the estimation of the magnetic field strength at the molecular C$_2$H$_2$ as an upper limit from the magnetic pressure and the thermal pressure forces:

\begin{equation}\label{magnfield}
\frac{B^2}{8~\pi} \leq n_\text{gas} \cdot k_b \cdot T,
\end{equation}

from which we obtain the estimate of $B\approx7.6$~G, where $k_B$ is the Boltzmann constant. However, considering the conservative estimation of the magnetic field ($\sim~10~$G), the corresponding Alfv\'{e}n speed can be then estimated as:

\begin{equation}
v_\text{A} = \frac{B}{\sqrt{4 \pi \rho_{\text{ionized}}}},
\end{equation}


where $\rho_{\text{ionized}}$ is the gas mass density of the ionized material, and $B$ is the magnetic field. With the gas mass density of the ionized material given above, we obtain the following estimations of the Alfv\'{e}n velocities: $v_\text{A} \approx410$~km/s, $<570$~km/s, and $<360$~km/s, using respectively the ionization fraction from \cite{groenewegen97}, \cite{spergel83}, and \cite{drake91}.


While these calculations do not prove the presence of Alfv\'{e}n waves in V~Oph, or that the variation in the C$_2$H$_2$ shell is related to this mechanism, they offer upper limits for the Alfv\'{e}n waves velocity, showing that it could be a reasonable possibility in the presence of a material with a higher degree of ionization, or higher gas mass. Indeed, if the mass density of the ionized material would be even a factor of $\sim100$ higher, the velocity would be lower, matching the observed $\sim20$~km/s inferred for the C2H2 shell variability. In the same direction, if the ionization factor would be higher, the Alfv\'{e}n wave velocity value would decrease to match the observations. Future detailed MHD calculations, which go beyond the purpose of this paper, would give further constraints on the Alfv\'{e}n waves velocity.

Alfv\'{e}n waves occur in the weakly ionized environment of the solar photosphere \citep{vranjes08}. Alfv\'{e}n waves have been found to be efficient in driving winds and produce clumpy mass loss from giant and supergiant stars (e.g.~\citealp{vlemmings14}; see also \citealp{hartmanmacgregor, airapetian00, suzuki06, airapetian10, airapetian15, yasuda19, carpenter18}, and Figure~$7$ in \citealp{rau18}). In a weakly ionized but collisional shell, a propagating Alfv\'{e}n wave would also involve the motion of the neutral species that are present in the shell. This is due to the friction between charged particles and neutrals (see e.g.~\citealp{tanenbaum}, \citealp{woods}, \citealp{jephcott62}, and many subsequent works, e.g., \citealp{kulsrud69}, \citealp{pudritz90}, \citealp{haerendel92}, \citealp{depontieu98}, \citealp{watts04}). This is valid for any weakly ionized plasma, including the lower solar atmosphere and extended atmospheric environments of cool stars. The shell ionization could be caused by the heating from shocks and galactic cosmic rays (\citealp{hoefnerandolofsson18}, \citealp{harper13}, \citealp{wedemeyer17}).

\subsubsection{Pattern motion}

A further hypothesis could be related to a matter of molecules appearing at a certain distance from the star because of favorable conditions at that epoch. This would generate a pattern speed, rather than material moving at a high velocity. In this case the observed time variations would not be related to the motion of material but to a change in the region of molecules/dust formation. This could cause the molecules to be formed at one epoch, then gets destroyed, and at the second epoch to form at another distance, then gets destroyed again, and so on. Therefore, molecules would not physically move, but just get formed and destroyed at different radii, at different epochs.



\subsection{An hypothesis of V~Oph reclassification to semi-regular star}\label{voph_semireg_vs_mira}

The period of V~Oph is $297$~d \citep{GCVS}, which is relatively short for Mira stars, and similar to the period of e.g.~the semi-regular star \object{Y~Pav} (see Table~1 in \citealp{rau17}). Not surprisingly indeed, also the observed visibilities vs.~wavelength at the shortest baseline $41.9$~m (see middle panel, left, Fig.~\ref{fig_interf_voph}) show a profile similar to semi-regular, rather then to Mira stars. We observed the same behavior in the visibility vs.~baselines (see Figures~\ref{interf-voph-visbase1}, \ref{interf-voph-visbase2}, \ref{interf-voph-visbase3}), in terms of radial extension of the models. 

To verify this hypothesis, following \cite{wittkowski17} we placed V~Oph in the period-luminosity (P-L) diagram by \cite{wood15} (Fig.~$10$), where pulsation sequences are designated. Even though the diagram is based on the Large Magellanic Cloud (LMC), \cite{whitelock08} did not find significant differences in the P-L relations at different metallicities. Correcting the 2MASS catalogue J and K magnitudes \citep{cutri03} for the LMC distance, we calculated the W$_{\text{JK}}$ index value of V~Oph: W$_{JK} = 9.37~\pm~0.3$. (W$_{\text{JK}} =$ K - $0.686$(J - K) is a reddening-free measure of the luminosity, -- cf.~Fig.~$1$ in~\citealp{wood15}). Considering its pulsation period, V~Oph is located on the sequence that corresponds to the radial first overtone mode pulsation, which is more similar to semi-regular stars, rather than Mira stars. In addition, from the W$_{\text{JK}}$ relation, V~Oph mass is estimated to be between $1.6$ and $2.4$~M$_\odot$ (see Fig.~$8$ in \citealp{wood15}).

We could interpret these results as V~Oph being more similar to the behavior of a semi-regular star, rather than a Mira. While \cite{feast87} suggested that semi-regulars will become Miras as the stars evolve, and \cite{habing96} instead claimed the opposite, other authors (e.g.~\citealp{kerschbaum92, speck00}, and references therein) advocate that the evolution of Mira and Semi-regular stars is likely to be non-monotonic, with the stars alternating between being Mira and SRs.


\section{Conclusions}\label{conclu}

We have compared archive VLTI/MIDI observations of the carbon-rich star V~Oph with the DARWIN grid of self-consistent dynamic atmosphere models for carbon stars. 

The photometric modeling agrees extremely well with the literature data, but we underline that spectrum of V~Oph covering wavelength from the visual to the mid-infrared is not available in the literature. Spectral observations with the NASA Infrared Telescope Facility (IRTF) SpeX and BASS instruments covering a wide wavelength range would be of extreme importance for future modeling efforts.

Our results from the interferometric modeling exhibit a fair agreement with the observations with some discrepancies. The parameters of the molecular outer atmosphere and the dust shell derived using the self-consistent dynamical models agree with those derived based on the semi-empirical modeling by OH07, given the different modeling approaches, showing that the star radius is bigger at minimum light, i.e.~at $\phi = 0.49$, than at phases $0.18$ and $0.65$. We calculate the photospheric size through our modeling: respectively $479$, $494$, $448$~R$_{\odot}$ at the three phases; and the dust radii: $780$, $853$, $787$~R$_{\odot}$. 

The interferometric modeling of V~Oph shows that this star may be more similar to a semi-regular star, rather than a Mira (e.g., the compact atmosphere, and the comparison with previous works); we further confirm this hypothesis using the pulsation sequences and relation in \cite{wood15}, which shows that V~Oph might pulsate in the first overtone mode rather than the fundamental mode.

To explain the interferometric variability at the different phases, we calculate the characteristic speed of perturbation. This time variation could be interpreted as a change in the density and temperature at different regions, or in the context of the Alfv\'{e}n waves. Such waves could propagate in a magnetized plasma, and we provide upper limits for the Alfv\'{e}n waves velocity. The strength of the magnetic field is: $B\sim7.6$~G. This value is in agreement with the findings in literature for AGB stars. Such Alfv\'{e}n waves should be detectable as a non-thermal broadening in cool UV lines during this transition. Detailed MHD modeling is needed for providing further constrains on the Alfv\'{e}n waves calculations. 


Simultaneous near- and mid-infrared interferometric (snapshot) monitoring observations of V~Oph with latest $2^\text{nd}$ generation VLTI instrument MATISSE \citep{lopez14} will be essential in order to better understand the phase and cycle dependence of the physical properties of the outer atmosphere and the dust shell. Finally, MATISSE will also be able to imaging this star, which will substantially help in studying the distribution and formation of molecular and dust environment of the C-rich star V~Oph. Consequently, these observations would help constraining temperature at different regions, and help to verify if the time variation is due to changes in the density and temperature at different regions, or if the Alfv\'{e}n wave-driven winds could play a role in explaining the extended molecular and dust layers.

\acknowledgments
We thank the anonymous referee for the useful comments, which helped improving the quality of the paper. We acknowledge the use of the Uppsala model provided by Prof.~Kjell Eriksson (private comm.). G.~R.~thanks Prof.~Josef Hron, his FWF project P23006, and his team in Vienna, for the fruitful discussions and support in the past years. G.~R.~also thanks ESO for supporting her 2018 visit at ESO/Garching that helped to further develop the discussion in this paper.
K.~O.~acknowledges the support of the Comisi\'{o}n Nacional de Investigaci\'{o}n Cient\'{i}fica y Tecnol\'{o}gica (CONCYT) through the FONDECYT Regular grant 1180066.
This work has made use of data from the European Space Agency (ESA) mission {\it Gaia} (\url{https://www.cosmos.esa.int/gaia}), processed by the {\it Gaia} Data Processing and Analysis Consortium (DPAC, \url{https://www.cosmos.esa.int/web/gaia/dpac/consortium}). Funding for the DPAC has been provided by national institutions, in particular the institutions participating in the {\it Gaia} Multilateral Agreement. 

\software{COMA} code \citep{aringer00, aringer09}
\software{{\AE}SOPUS} code \citep{marigoaringer09}

\bibliographystyle{apj}
\bibliography{master_papermugamma}

\begin{thebibliography}{}
\expandafter\ifx\csname natexlab\endcsname\relax\def\natexlab#1{#1}\fi

\bibitem[{{Airapetian} {et~al.}(2010){Airapetian}, {Carpenter}, \&
  {Ofman}}]{airapetian10}
{Airapetian}, V., {Carpenter}, K.~G., \& {Ofman}, L. 2010, \apj, 723, 1210

\bibitem[{{Airapetian} {et~al.}(2015){Airapetian}, {Leake}, \&
  {Carpenter}}]{airapetian15}
{Airapetian}, V., {Leake}, J., \& {Carpenter}, K. 2015, IAU General Assembly,
  21, 2190977

\bibitem[{{Airapetian} {et~al.}(2000){Airapetian}, {Ofman}, {Robinson},
  {Carpenter}, \& {Davila}}]{airapetian00}
{Airapetian}, V.~S., {Ofman}, L., {Robinson}, R.~D., {Carpenter}, K., \&
  {Davila}, J. 2000, \apj, 528, 965

\bibitem[{{Alfonso-Garz{\'o}n} {et~al.}(2012){Alfonso-Garz{\'o}n}, {Domingo},
  {Mas-Hesse}, \& {Gim{\'e}nez}}]{alf12}
{Alfonso-Garz{\'o}n}, J., {Domingo}, A., {Mas-Hesse}, J.~M., \& {Gim{\'e}nez},
  A. 2012, \aap, 548, A79

\bibitem[{{Aringer}(2000)}]{aringer00}
{Aringer}, B. 2000, in IAU Symposium, Vol. 177, The Carbon Star Phenomenon, ed.
  R.~F. {Wing}, 519

\bibitem[{{Aringer} {et~al.}(2009){Aringer}, {Girardi}, {Nowotny}, {Marigo}, \&
  {Lederer}}]{aringer09}
{Aringer}, B., {Girardi}, L., {Nowotny}, W., {Marigo}, P., \& {Lederer}, M.~T.
  2009, \aap, 503, 913

\bibitem[{Bailer-Jones {et~al.}(2018)Bailer-Jones, Rybizki, Fouesneau,
  Mantelet, \& Andrae}]{BailerJones18}
Bailer-Jones, C. A.~L., Rybizki, J., Fouesneau, M., Mantelet, G., \& Andrae, R.
  2018, The Astronomical Journal, 156, 58

\bibitem[{Ballester {et~al.}(2018)Ballester, Alexeev, Collados, Downes, Pfaff,
  Gilbert, Khodachenko, Khomenko, Shaikhislamov, Soler, V{\'a}zquez-Semadeni,
  \& Zaqarashvili}]{ballester18}
Ballester, J.~L., Alexeev, I., Collados, M., {et~al.} 2018, Space Science
  Reviews, 214, 58

\bibitem[{{Bergeat} \& {Chevallier}(2005)}]{bergeat05}
{Bergeat}, J., \& {Chevallier}, L. 2005, \aap, 429, 235

\bibitem[{{Bladh} {et~al.}(2019{\natexlab{a}}){Bladh}, {Eriksson}, {Marigo},
  {Liljegren}, \& {Aringer}}]{bladh19}
{Bladh}, S., {Eriksson}, K., {Marigo}, P., {Liljegren}, S., \& {Aringer}, B.
  2019{\natexlab{a}}, arXiv e-prints, arXiv:1902.05352

\bibitem[{{Bladh} {et~al.}(2019{\natexlab{b}}){Bladh}, {Liljegren},
  {H{\"o}fner}, {Aringer}, \& {Marigo}}]{bladh19b}
{Bladh}, S., {Liljegren}, S., {H{\"o}fner}, S., {Aringer}, B., \& {Marigo}, P.
  2019{\natexlab{b}}, arXiv e-prints, arXiv:1904.10943

\bibitem[{{Bressan} {et~al.}(2012){Bressan}, {Marigo}, {Girardi}, {Salasnich},
  {Dal Cero}, {Rubele}, \& {Nanni}}]{bressan12}
{Bressan}, A., {Marigo}, P., {Girardi}, L., {et~al.} 2012, \mnras, 427, 127

\bibitem[{Carpenter {et~al.}(2018)Carpenter, Nielsen, Kober, Ayres, Wahlgren,
  \& Rau}]{carpenter18}
Carpenter, K.~G., Nielsen, K.~E., Kober, G.~V., {et~al.} 2018, The
  Astrophysical Journal, 869, 157

\bibitem[{{Cernicharo} {et~al.}(2000){Cernicharo}, {Gu{\'e}lin}, \&
  {Kahane}}]{cernicharo00}
{Cernicharo}, J., {Gu{\'e}lin}, M., \& {Kahane}, C. 2000, \aaps, 142, 181

\bibitem[{{Cherchneff}(2012)}]{cherchneff12}
{Cherchneff}, I. 2012, \aap, 545, A12

\bibitem[{{Clayton}(2012)}]{clayton12}
{Clayton}, G.~C. 2012, Journal of the American Association of Variable Star
  Observers (JAAVSO), 40, 539

\bibitem[{{Cristallo} {et~al.}(2007){Cristallo}, {Straniero}, {Lederer}, \&
  {Aringer}}]{cristallo07}
{Cristallo}, S., {Straniero}, O., {Lederer}, M.~T., \& {Aringer}, B. 2007,
  \apj, 667, 489

\bibitem[{{Cruzal{\`e}bes} {et~al.}(2013){Cruzal{\`e}bes}, {Jorissen},
  {Rabbia}, {Sacuto}, {Chiavassa}, {Pasquato}, {Plez}, {Eriksson}, {Spang}, \&
  {Chesneau}}]{cruzalbes13}
{Cruzal{\`e}bes}, P., {Jorissen}, A., {Rabbia}, Y., {et~al.} 2013, \mnras, 434,
  437

\bibitem[{{Cutri} {et~al.}(2003{\natexlab{a}}){Cutri}, {Skrutskie}, {van Dyk},
  {Beichman}, {Carpenter}, {Chester}, {Cambresy}, {Evans}, {Fowler}, {Gizis},
  {Howard}, {Huchra}, {Jarrett}, {Kopan}, {Kirkpatrick}, {Light}, {Marsh},
  {McCallon}, {Schneider}, {Stiening}, {Sykes}, {Weinberg}, {Wheaton},
  {Wheelock}, \& {Zacarias}}]{cutri03}
{Cutri}, R.~M., {Skrutskie}, M.~F., {van Dyk}, S., {et~al.} 2003{\natexlab{a}},
  {2MASS All Sky Catalog of point sources.} (Caltech)

\bibitem[{{Cutri} {et~al.}(2003{\natexlab{b}}){Cutri}, {Skrutskie}, {van Dyk},
  {Beichman}, {Carpenter}, {Chester}, {Cambresy}, {Evans}, {Fowler}, {Gizis},
  {Howard}, {Huchra}, {Jarrett}, {Kopan}, {Kirkpatrick}, {Light}, {Marsh},
  {McCallon}, {Schneider}, {Stiening}, {Sykes}, {Weinberg}, {Wheaton},
  {Wheelock}, \& {Zacarias}}]{2mass}
---. 2003{\natexlab{b}}, VizieR Online Data Catalog, II/246

\bibitem[{{Davis} {et~al.}(2000){Davis}, {Tango}, \& {Booth}}]{davis00}
{Davis}, J., {Tango}, W.~J., \& {Booth}, A.~J. 2000, \mnras, 318, 387

\bibitem[{{de Pontieu} \& {Haerendel}(1998)}]{depontieu98}
{de Pontieu}, B., \& {Haerendel}, G. 1998, \aap, 338, 729

\bibitem[{{DENIS Consortium}(2005)}]{denis}
{DENIS Consortium}. 2005, VizieR Online Data Catalog, 2263

\bibitem[{{Drake} {et~al.}(1991){Drake}, {Linsky}, {Judge}, \&
  {Elitzur}}]{drake91}
{Drake}, S.~A., {Linsky}, J.~L., {Judge}, P.~G., \& {Elitzur}, M. 1991, \aj,
  101, 230

\bibitem[{{Duthu} {et~al.}(2017){Duthu}, {Herpin}, {Wiesemeyer}, {Baudry},
  {L{\`e}bre}, \& {Paubert}}]{duthu17}
{Duthu}, A., {Herpin}, F., {Wiesemeyer}, H., {et~al.} 2017, \aap, 604, A12

\bibitem[{{Eriksson} {et~al.}(2014){Eriksson}, {Nowotny}, {H{\"o}fner},
  {Aringer}, \& {Wachter}}]{erik14}
{Eriksson}, K., {Nowotny}, W., {H{\"o}fner}, S., {Aringer}, B., \& {Wachter},
  A. 2014, \aap, 566, A95

\bibitem[{{Falceta-Gon{\c{c}}alves} \&
  {Jatenco-Pereira}(2002)}]{falcetagoncalves02}
{Falceta-Gon{\c{c}}alves}, D., \& {Jatenco-Pereira}, V. 2002, \apj, 576, 976

\bibitem[{{Feast} \& {Whitelock}(1987)}]{feast87}
{Feast}, M.~W., \& {Whitelock}, P.~A. 1987, in Astrophysics and Space Science
  Library, Vol. 132, Late Stages of Stellar Evolution, ed. S.~{Kwok} \& S.~R.
  {Pottasch}, 33--46

\bibitem[{{Fleischer} {et~al.}(1992){Fleischer}, {Gauger}, \&
  {Sedlmayr}}]{Fleischer92}
{Fleischer}, A.~J., {Gauger}, A., \& {Sedlmayr}, E. 1992, \aap, 266, 321

\bibitem[{{Gaia Collaboration} {et~al.}(2018){Gaia Collaboration}, {Brown},
  {Vallenari}, {Prusti}, {de Bruijne}, {Babusiaux}, \&
  {Bailer-Jones}}]{gaia2018}
{Gaia Collaboration}, {Brown}, A.~G.~A., {Vallenari}, A., {et~al.} 2018, ArXiv
  e-prints, arXiv:1804.09365

\bibitem[{{Gaia Collaboration} {et~al.}(2016){Gaia Collaboration}, {Prusti},
  {de Bruijne}, {Brown}, {Vallenari}, {Babusiaux}, {Bailer-Jones}, {Bastian},
  {Biermann}, {Evans}, \& et~al.}]{gaia2016}
{Gaia Collaboration}, {Prusti}, T., {de Bruijne}, J.~H.~J., {et~al.} 2016,
  \aap, 595, A1

\bibitem[{{Gail} \& {Sedlmayr}(1988)}]{gail88}
{Gail}, H.-P., \& {Sedlmayr}, E. 1988, \aap, 206, 153

\bibitem[{{Gauger} {et~al.}(1990){Gauger}, {Sedlmayr}, \& {Gail}}]{gauger90}
{Gauger}, A., {Sedlmayr}, E., \& {Gail}, H.-P. 1990, \aap, 235, 345

\bibitem[{{Gautschy-Loidl} {et~al.}(2004){Gautschy-Loidl}, {H{\"o}fner},
  {J{\o}rgensen}, \& {Hron}}]{Gautschy-Loidl04}
{Gautschy-Loidl}, R., {H{\"o}fner}, S., {J{\o}rgensen}, U.~G., \& {Hron}, J.
  2004, \aap, 422, 289

\bibitem[{{Gong} {et~al.}(2015){Gong}, {Henkel}, {Spezzano}, {Thorwirth},
  {Menten}, {Wyrowski}, {Mao}, \& {Klein}}]{gong15}
{Gong}, Y., {Henkel}, C., {Spezzano}, S., {et~al.} 2015, \aap, 574, A56

\bibitem[{{Groenewegen}(1997)}]{groenewegen97}
{Groenewegen}, M.~A.~T. 1997, \aap, 317, 503

\bibitem[{{Groenewegen} {et~al.}(1999){Groenewegen}, {Baas}, {Blommaert},
  {Stehle}, {Josselin}, \& {Tilanus}}]{groenewegen99}
{Groenewegen}, M.~A.~T., {Baas}, F., {Blommaert}, J.~A.~D.~L., {et~al.} 1999,
  \aaps, 140, 197

\bibitem[{{Gunther}(2006)}]{afoev}
{Gunther}, J. 2006, Journal of the American Association of Variable Star
  Observers (JAAVSO), 35, 208

\bibitem[{{Habing}(1996)}]{habing96}
{Habing}, H.~J. 1996, \aapr, 7, 97

\bibitem[{{Haerendel}(1992)}]{haerendel92}
{Haerendel}, G. 1992, \nat, 360, 241

\bibitem[{Harper {et~al.}(2012)Harper, O'Riain, \& Ayres}]{harper13}
Harper, G.~M., O'Riain, N., \& Ayres, T.~R. 2012, Monthly Notices of the Royal
  Astronomical Society, 428, 2064

\bibitem[{{Hartmann} \& {MacGregor}(1982)}]{hartmanmacgregor}
{Hartmann}, L., \& {MacGregor}, K.~B. 1982, \apj, 257, 264

\bibitem[{{Henden} {et~al.}(2016){Henden}, {Templeton}, {Terrell}, {Smith},
  {Levine}, \& {Welch}}]{aavso}
{Henden}, A.~A., {Templeton}, M., {Terrell}, D., {et~al.} 2016, VizieR Online
  Data Catalog, 2336

\bibitem[{{Hinkle}(1978)}]{hinkle78}
{Hinkle}, K.~H. 1978, \apj, 220, 210

\bibitem[{{H{\"o}fner}(1999)}]{hofner99}
{H{\"o}fner}, S. 1999, \aap, 346, L9

\bibitem[{{H{\"o}fner}(2007)}]{hofner07}
{H{\"o}fner}, S. 2007, in Astronomical Society of the Pacific Conference
  Series, Vol. 378, Why Galaxies Care About AGB Stars: Their Importance as
  Actors and Probes, ed. F.~{Kerschbaum}, C.~{Charbonnel}, \& R.~F. {Wing}, 145

\bibitem[{{H{\"o}fner} {et~al.}(2016){H{\"o}fner}, {Bladh}, {Aringer}, \&
  {Ahuja}}]{hofner16}
{H{\"o}fner}, S., {Bladh}, S., {Aringer}, B., \& {Ahuja}, R. 2016, \aap, 594,
  A108

\bibitem[{{H{\"o}fner} \& {Dorfi}(1997)}]{HoefnerDorfi}
{H{\"o}fner}, S., \& {Dorfi}, E.~A. 1997, \aap, 319, 648

\bibitem[{{H{\"o}fner} \& {Freytag}(2019)}]{hoefnerandfreytag19}
{H{\"o}fner}, S., \& {Freytag}, B. 2019, arXiv e-prints, arXiv:1902.04074

\bibitem[{{H{\"o}fner} {et~al.}(2003){H{\"o}fner}, {Gautschy-Loidl}, {Aringer},
  \& {J{\o}rgensen}}]{hofner03}
{H{\"o}fner}, S., {Gautschy-Loidl}, R., {Aringer}, B., \& {J{\o}rgensen}, U.~G.
  2003, \aap, 399, 589

\bibitem[{{H{\"o}fner} \& {Olofsson}(2018)}]{hoefnerandolofsson18}
{H{\"o}fner}, S., \& {Olofsson}, H. 2018, \aapr, 26, 1

\bibitem[{{Iben} \& {Renzini}(1983)}]{ibenrenzini83}
{Iben}, Jr., I., \& {Renzini}, A. 1983, \araa, 21, 271

\bibitem[{{Jatenco-Pereira}(2014)}]{jatencopereira14}
{Jatenco-Pereira}, V. 2014, in Revista Mexicana de Astronomia y Astrofisica
  Conference Series, Vol.~44, 140--140

\bibitem[{Jephcott(1962)}]{jephcott62}
Jephcott, D.~F.~and~Stocker, P.~M. 1962, Fluid Mech., 13, 587

\bibitem[{{Katz} {et~al.}(2019){Katz}, {Sartoretti}, {Cropper}, {Panuzzo},
  {Seabroke}, {Viala}, {Benson}, {Blomme}, {Jasniewicz}, {Jean-Antoine},
  {Huckle}, {Smith}, {Baker}, {Crifo}, {Damerdji}, {David}, {Dolding},
  {Fr{\'e}mat}, {Gosset}, {Guerrier}, {Guy}, {Haigron}, {Jan{\ss}en},
  {Marchal}, {Plum}, {Soubiran}, {Th{\'e}venin}, {Ajaj}, {Allende Prieto},
  {Babusiaux}, {Boudreault}, {Chemin}, {Delle Luche}, {Fabre}, {Gueguen},
  {Hambly}, {Lasne}, {Meynadier}, {Pailler}, {Panem}, {Royer}, {Tauran},
  {Zurbach}, {Zwitter}, {Arenou}, {Bossini}, {Gerssen}, {G{\'o}mez},
  {Lemaitre}, {Leclerc}, {Morel}, {Munari}, {Turon}, {Vallenari}, \&
  {{\v{Z}}erjal}}]{katz19}
{Katz}, D., {Sartoretti}, P., {Cropper}, M., {et~al.} 2019, \aap, 622, A205

\bibitem[{{Kerschbaum} \& {Hron}(1992)}]{kerschbaum92}
{Kerschbaum}, F., \& {Hron}, J. 1992, \aap, 263, 97

\bibitem[{{Klotz} {et~al.}(2013){Klotz}, {Paladini}, {Hron}, {Aringer},
  {Sacuto}, {Marigo}, \& {Verhoelst}}]{klotz13}
{Klotz}, D., {Paladini}, C., {Hron}, J., {et~al.} 2013, \aap, 550, A86

\bibitem[{Kulsrud(1969)}]{kulsrud69}
Kulsrud, R.~and~Pierce, W. 1969, \apj, 156, 445

\bibitem[{{Leinert} {et~al.}(2003){Leinert}, {Graser}, {Richichi},
  {Sch{\"o}ller}, {Waters}, {Perrin}, {Jaffe}, {Lopez}, {Glazenborg-Kluttig},
  {Przygodda}, {Morel}, {Biereichel}, {Haddad}, {Housen}, \&
  {Wallander}}]{midi2003}
{Leinert}, C., {Graser}, U., {Richichi}, A., {et~al.} 2003, The Messenger, 112,
  13

\bibitem[{{Loidl}(2001)}]{loidl_phdthesis}
{Loidl}, R. 2001, PhD thesis, Institute for Astrophysics , University of
  Vienna, Austria

\bibitem[{{Loidl} {et~al.}(1999){Loidl}, {H{\"o}fner}, {J{\o}rgensen}, \&
  {Aringer}}]{loidl99}
{Loidl}, R., {H{\"o}fner}, S., {J{\o}rgensen}, U.~G., \& {Aringer}, B. 1999,
  \aap, 342, 531

\bibitem[{{Loidl} {et~al.}(2001){Loidl}, {Lan{\c c}on}, \&
  {J{\o}rgensen}}]{loidl01}
{Loidl}, R., {Lan{\c c}on}, A., \& {J{\o}rgensen}, U.~G. 2001, \aap, 371, 1065

\bibitem[{{Lopez} {et~al.}(2014){Lopez}, {Lagarde}, {Jaffe}, {Petrov},
  {Sch{\"o}ller}, {Antonelli}, {Beckmann}, {Berio}, {Bettonvil}, {Glindemann},
  {Gonzalez}, {Graser}, {Hofmann}, {Millour}, {Robbe-Dubois}, {Venema}, {Wolf},
  {Henning}, {Lanz}, {Weigelt}, {Agocs}, {Bailet}, {Bresson}, {Bristow},
  {Dugu{\'e}}, {Heininger}, {Kroes}, {Laun}, {Lehmitz}, {Neumann}, {Augereau},
  {Avila}, {Behrend}, {van Belle}, {Berger}, {van Boekel}, {Bonhomme},
  {Bourget}, {Brast}, {Clausse}, {Connot}, {Conzelmann}, {Cruzal{\`e}bes},
  {Csepany}, {Danchi}, {Delbo}, {Delplancke}, {Dominik}, {van Duin}, {Elswijk},
  {Fantei}, {Finger}, {Gabasch}, {Gay}, {Girard}, {Girault}, {Gitton},
  {Glazenborg}, {Gont{\'e}}, {Guitton}, {Guniat}, {De Haan}, {Haguenauer},
  {Hanenburg}, {Hogerheijde}, {ter Horst}, {Hron}, {Hugues}, {Hummel},
  {Idserda}, {Ives}, {Jakob}, {Jasko}, {Jolley}, {Kiraly}, {K{\"o}hler},
  {Kragt}, {Kroener}, {Kuindersma}, {Labadie}, {Leinert}, {Le Poole}, {Lizon},
  {Lucuix}, {Marcotto}, {Martinache}, {Martinot-Lagarde}, {Mathar}, {Matter},
  {Mauclert}, {Mehrgan}, {Meilland}, {Meisenheimer}, {Meisner}, {Mellein},
  {Menardi}, {Menut}, {Merand}, {Morel}, {Mosoni}, {Navarro}, {Nussbaum},
  {Ottogalli}, {Palsa}, {Panduro}, {Pantin}, {Parra}, {Percheron}, {Duc},
  {Pott}, {Pozna}, {Przygodda}, {Rabbia}, {Richichi}, {Rigal}, {Roelfsema},
  {Rupprecht}, {Schertl}, {Schmidt}, {Schuhler}, {Schuil}, {Spang},
  {Stegmeier}, {Thiam}, {Tromp}, {Vakili}, {Vannier}, {Wagner}, \&
  {Woillez}}]{lopez14}
{Lopez}, B., {Lagarde}, S., {Jaffe}, W., {et~al.} 2014, The Messenger, 157, 5

\bibitem[{{Luri} {et~al.}(2018){Luri}, {Brown}, {Sarro}, {Arenou},
  {Bailer-Jones}, {Castro-Ginard}, {de Bruijne}, {Prusti}, {Babusiaux}, \&
  {Delgado}}]{gaiaparallaxes}
{Luri}, X., {Brown}, A.~G.~A., {Sarro}, L.~M., {et~al.} 2018, ArXiv e-prints,
  arXiv:1804.09376

\bibitem[{{Marigo} \& {Aringer}(2009)}]{marigoaringer09}
{Marigo}, P., \& {Aringer}, B. 2009, \aap, 508, 1539

\bibitem[{{Marigo} {et~al.}(2013){Marigo}, {Bressan}, {Nanni}, {Girardi}, \&
  {Pumo}}]{marigo13}
{Marigo}, P., {Bressan}, A., {Nanni}, A., {Girardi}, L., \& {Pumo}, M.~L. 2013,
  \mnras, 434, 488

\bibitem[{{Mattsson} {et~al.}(2010){Mattsson}, {Wahlin}, \&
  {H{\"o}fner}}]{mattsson10}
{Mattsson}, L., {Wahlin}, R., \& {H{\"o}fner}, S. 2010, \aap, 509, A14

\bibitem[{{McDonald} {et~al.}(2017){McDonald}, {Zijlstra}, \&
  {Watson}}]{mcdonald17}
{McDonald}, I., {Zijlstra}, A.~A., \& {Watson}, R.~A. 2017, \mnras, 471, 770

\bibitem[{{McKee} \& {Zweibel}(1995)}]{mckee95}
{McKee}, C.~F., \& {Zweibel}, E.~G. 1995, \apj, 440, 686

\bibitem[{{Nowotny} {et~al.}(2013){Nowotny}, {Aringer}, {H{\"o}fner}, \&
  {Eriksson}}]{nowotny13}
{Nowotny}, W., {Aringer}, B., {H{\"o}fner}, S., \& {Eriksson}, K. 2013, \aap,
  552, A20

\bibitem[{{Nowotny} {et~al.}(2011){Nowotny}, {Aringer}, {H{\"o}fner}, \&
  {Lederer}}]{walter2}
{Nowotny}, W., {Aringer}, B., {H{\"o}fner}, S., \& {Lederer}, M.~T. 2011, \aap,
  529, A129

\bibitem[{{Nowotny} {et~al.}(2010){Nowotny}, {H{\"o}fner}, \&
  {Aringer}}]{nowotny10}
{Nowotny}, W., {H{\"o}fner}, S., \& {Aringer}, B. 2010, \aap, 514, A35

\bibitem[{{Nowotny} {et~al.}(2005){Nowotny}, {Lebzelter}, {Hron}, \&
  {H{\"o}fner}}]{nowotny05}
{Nowotny}, W., {Lebzelter}, T., {Hron}, J., \& {H{\"o}fner}, S. 2005, \aap,
  437, 285

\bibitem[{{Ohnaka} {et~al.}(2007){Ohnaka}, {Driebe}, {Weigelt}, \&
  {Wittkowski}}]{ohnaka07}
{Ohnaka}, K., {Driebe}, T., {Weigelt}, G., \& {Wittkowski}, M. 2007, \aap, 466,
  1099

\bibitem[{{Olnon} {et~al.}(1986){Olnon}, {Raimond}, {Neugebauer}, {van Duinen},
  {Habing}, {Aumann}, {Beintema}, {Boggess}, {Borgman}, {Clegg}, {Gillett},
  {Hauser}, {Houck}, {Jennings}, {de Jong}, {Low}, {Marsden}, {Pottasch},
  {Soifer}, {Walker}, {Emerson}, {Rowan-Robinson}, {Wesselius}, {Baud},
  {Beichman}, {Gautier}, {Harris}, {Miley}, \& {Young}}]{iras-satellite}
{Olnon}, F.~M., {Raimond}, E., {Neugebauer}, G., {et~al.} 1986, \aaps, 65, 607

\bibitem[{{Olofsson} {et~al.}(1993){Olofsson}, {Eriksson}, {Gustafsson}, \&
  {Carlstroem}}]{olofsson93}
{Olofsson}, H., {Eriksson}, K., {Gustafsson}, B., \& {Carlstroem}, U. 1993,
  \apjs, 87, 305

\bibitem[{{Pegourie}(1988)}]{pegourie}
{Pegourie}, B. 1988, \aap, 194, 335

\bibitem[{{Pojmanski}(2002)}]{asas}
{Pojmanski}, G. 2002, \actaa, 52, 397

\bibitem[{{Pudritz}(1990)}]{pudritz90}
{Pudritz}, R.~E. 1990, \apj, 350, 195

\bibitem[{{Rau}(2016)}]{rau16}
{Rau}, G. 2016, PhD thesis, University of Vienna - Institute for Astrophysics

\bibitem[{{Rau} {et~al.}(2017){Rau}, {Hron}, {Paladini}, {Aringer}, {Eriksson},
  {Marigo}, {Nowotny}, \& {Grellmann}}]{rau17}
{Rau}, G., {Hron}, J., {Paladini}, C., {et~al.} 2017, \aap, 600, A92

\bibitem[{{Rau} {et~al.}(2018){Rau}, {Nielsen}, {Carpenter}, \&
  {Airapetian}}]{rau18}
{Rau}, G., {Nielsen}, K.~E., {Carpenter}, K.~G., \& {Airapetian}, V. 2018,
  \apj, 869, 1

\bibitem[{{Rau} {et~al.}(2015){Rau}, {Paladini}, {Hron}, {Aringer},
  {Groenewegen}, \& {Nowotny}}]{rau15}
{Rau}, G., {Paladini}, C., {Hron}, J., {et~al.} 2015, \aap, 583, A106

\bibitem[{{Rau} {et~al.}(2019){Rau}, {Montez}, {Carpenter}, {Wittkowski},
  {Bladh}, {Karovska}, {Airapetian}, {Ayres}, {Boyer}, {Chiavassa}, {Clayton},
  {Danchi}, {De Marco}, {Dupree}, {Kaminski}, {Kastner}, {Kerschbaum},
  {Linsky}, {Lopez}, {Monnier}, {Montarg{\`e}s}, {Nielsen}, {Ohnaka},
  {Ramstedt}, {Roettenbacher}, {ten Brummelaar}, {Sarangi}, {van Belle}, \&
  {Ventura}}]{rau19}
{Rau}, G., {Montez}, Rodolfo, J., {Carpenter}, K., {et~al.} 2019, in \baas,
  Vol.~51, 241

\bibitem[{{Rouleau} \& {Martin}(1991)}]{roleau_and_martin}
{Rouleau}, F., \& {Martin}, P.~G. 1991, \apj, 377, 526

\bibitem[{{Sabin} {et~al.}(2015){Sabin}, {Wade}, \& {L{\`e}bre}}]{sabin15}
{Sabin}, L., {Wade}, G.~A., \& {L{\`e}bre}, A. 2015, \mnras, 446, 1988

\bibitem[{{Sacuto} {et~al.}(2011){Sacuto}, {Aringer}, {Hron}, {Nowotny},
  {Paladini}, {Verhoelst}, \& {H{\"o}fner}}]{sacuto11}
{Sacuto}, S., {Aringer}, B., {Hron}, J., {et~al.} 2011, \aap, 525, A42

\bibitem[{{Samus} {et~al.}(2009){Samus}, {Kazarovets}, {Pastukhova},
  {Tsvetkova}, \& {Durlevich}}]{GCVS}
{Samus}, N.~N., {Kazarovets}, E.~V., {Pastukhova}, E.~N., {Tsvetkova}, T.~M.,
  \& {Durlevich}, O.~V. 2009, \pasp, 121, 1378

\bibitem[{{Smith} {et~al.}(2004){Smith}, {Price}, \& {Baker}}]{dirbe}
{Smith}, B.~J., {Price}, S.~D., \& {Baker}, R.~I. 2004, \apjs, 154, 673

\bibitem[{{Speck} {et~al.}(2000){Speck}, {Barlow}, {Sylvester}, \&
  {Hofmeister}}]{speck00}
{Speck}, A.~K., {Barlow}, M.~J., {Sylvester}, R.~J., \& {Hofmeister}, A.~M.
  2000, \aaps, 146, 437

\bibitem[{{Spergel} {et~al.}(1983){Spergel}, {Giuliani}, \&
  {Knapp}}]{spergel83}
{Spergel}, D.~N., {Giuliani}, Jr., J.~L., \& {Knapp}, G.~R. 1983, \apj, 275,
  330

\bibitem[{{Suzuki}(2007)}]{suzuki06}
{Suzuki}, T.~K. 2007, \apj, 659, 1592

\bibitem[{Tanenbaum(1962)}]{tanenbaum}
Tanenbaum, S.~and~Mintzer, D. 1962, Physics of Fluids, 5, 1226

\bibitem[{{Tango} \& {Davis}(2002)}]{tangoanddavis02}
{Tango}, W.~J., \& {Davis}, J. 2002, \mnras, 333, 642

\bibitem[{{van Belle} {et~al.}(2013){van Belle}, {Paladini}, {Aringer}, {Hron},
  \& {Ciardi}}]{vanbelle13}
{van Belle}, G.~T., {Paladini}, C., {Aringer}, B., {Hron}, J., \& {Ciardi}, D.
  2013, \apj, 775, 45

\bibitem[{{Vidotto} \& {Jatenco-Pereira}(2006)}]{vidotto06}
{Vidotto}, A.~A., \& {Jatenco-Pereira}, V. 2006, \apj, 639, 416

\bibitem[{{Vlemmings} {et~al.}(2017){Vlemmings}, {Khouri}, {O'Gorman}, {De
  Beck}, {Humphreys}, {Lankhaar}, {Maercker}, {Olofsson}, {Ramstedt}, {Tafoya},
  \& {Takigawa}}]{vlemmings17}
{Vlemmings}, W., {Khouri}, T., {O'Gorman}, E., {et~al.} 2017, Nature Astronomy,
  1, 848

\bibitem[{{Vlemmings}(2014)}]{vlemmings14}
{Vlemmings}, W.~H.~T. 2014, in IAU Symposium, Vol. 302, Magnetic Fields
  throughout Stellar Evolution, ed. P.~{Petit}, M.~{Jardine}, \& H.~C.
  {Spruit}, 389--397

\bibitem[{{Vlemmings}(2018)}]{vlemmings18}
{Vlemmings}, W.~H.~T. 2018, Contributions of the Astronomical Observatory
  Skalnate Pleso, 48, 187

\bibitem[{{Vlemmings} {et~al.}(2018){Vlemmings}, {Khouri}, {De Beck},
  {Olofsson}, {Garc{\'\i}a-Segura}, {Villaver}, {Baudry}, {Humphreys},
  {Maercker}, \& {Ramstedt}}]{vlemmings18b}
{Vlemmings}, W.~H.~T., {Khouri}, T., {De Beck}, E., {et~al.} 2018, \aap, 613,
  L4

\bibitem[{{Vranjes} {et~al.}(2008){Vranjes}, {Poedts}, {Pandey}, \& {de
  Pontieu}}]{vranjes08}
{Vranjes}, J., {Poedts}, S., {Pandey}, B.~P., \& {de Pontieu}, B. 2008, \aap,
  478, 553

\bibitem[{{Watts} \& {Hanna}(2004)}]{watts04}
{Watts}, C., \& {Hanna}, J. 2004, Phys. Plasmas, 11, 1358

\bibitem[{{Wedemeyer} {et~al.}(2017){Wedemeyer}, {Ku{\v{c}}inskas}, {Klevas},
  \& {Ludwig}}]{wedemeyer17}
{Wedemeyer}, S., {Ku{\v{c}}inskas}, A., {Klevas}, J., \& {Ludwig}, H.-G. 2017,
  \aap, 606, A26

\bibitem[{{Whitelock} {et~al.}(2006){Whitelock}, {Feast}, {Marang}, \&
  {Groenewegen}}]{whitelock06}
{Whitelock}, P.~A., {Feast}, M.~W., {Marang}, F., \& {Groenewegen}, M.~A.~T.
  2006, \mnras, 369, 751

\bibitem[{{Whitelock} {et~al.}(2008){Whitelock}, {Feast}, \& {Van
  Leeuwen}}]{whitelock08}
{Whitelock}, P.~A., {Feast}, M.~W., \& {Van Leeuwen}, F. 2008, \mnras, 386, 313

\bibitem[{{Wittkowski} {et~al.}(2017){Wittkowski}, {Hofmann}, {H{\"o}fner}, {Le
  Bouquin}, {Nowotny}, {Paladini}, {Young}, {Berger}, {Brunner}, {de
  Gregorio-Monsalvo}, {Eriksson}, {Hron}, {Humphreys}, {Lindqvist}, {Maercker},
  {Mohamed}, {Olofsson}, {Ramstedt}, \& {Weigelt}}]{wittkowski17}
{Wittkowski}, M., {Hofmann}, K.~H., {H{\"o}fner}, S., {et~al.} 2017, \aap, 601,
  A3

\bibitem[{{Wittkowski} {et~al.}(2018){Wittkowski}, {Rau}, {Chiavassa},
  {H{\"o}fner}, {Scholz}, {Wood}, {de Wit}, {Eisenhauer}, {Haubois}, \&
  {Paumard}}]{wittkowski18}
{Wittkowski}, M., {Rau}, G., {Chiavassa}, A., {et~al.} 2018, \aap, 613, L7

\bibitem[{{Woitke}(2006)}]{woitke06}
{Woitke}, P. 2006, \aap, 460, L9

\bibitem[{{Wood}(2015)}]{wood15}
{Wood}, P.~R. 2015, \mnras, 448, 3829

\bibitem[{Woods(1962)}]{woods}
Woods. 1962, Phys. Review, 128, 1099

\bibitem[{{Yamamura} \& {de Jong}(2000)}]{yamamura00}
{Yamamura}, I., \& {de Jong}, T. 2000, in ESA Special Publication, Vol. 456,
  ISO Beyond the Peaks: The 2nd ISO Workshop on Analytical Spectroscopy, ed.
  A.~{Salama}, M.~F. {Kessler}, K.~{Leech}, \& B.~{Schulz}, 155

\bibitem[{{Yasuda} {et~al.}(2019){Yasuda}, {Suzuki}, \& {Kozasa}}]{yasuda19}
{Yasuda}, Y., {Suzuki}, T.~K., \& {Kozasa}, T. 2019, arXiv e-prints,
  arXiv:1905.09155

\end{thebibliography}

\clearpage
\begin{appendices}
\section{Windless model}\label{appendix_windless}

\begin{figure}[!htbp]
\begin{center}
\includegraphics[width=0.7\textwidth, bb=36 7 695 761, angle=0]{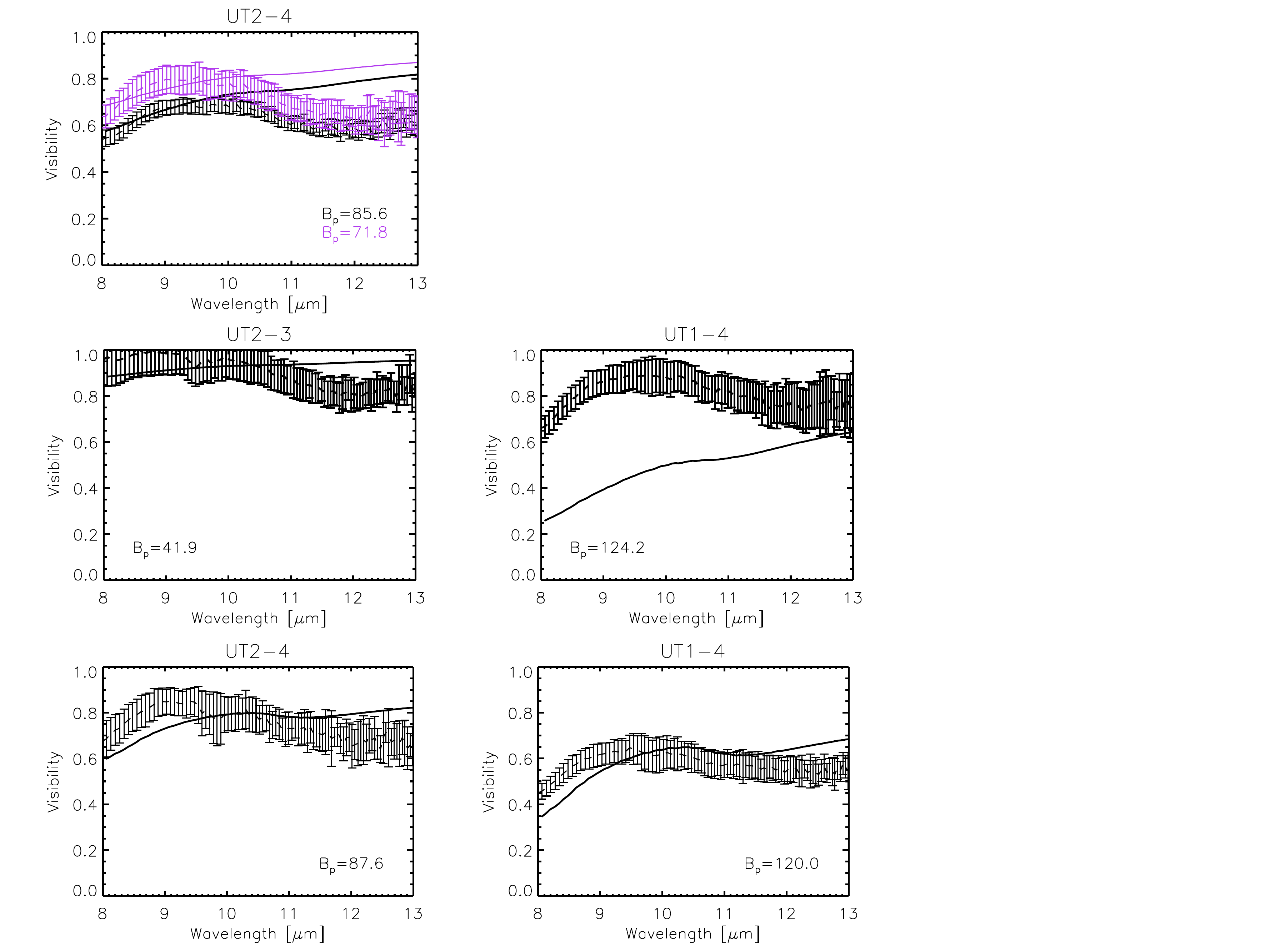}
\caption{As Fig.~\ref{fig_interf_voph}, but for the \textit{windless} model.}
\label{visib_windless}
\end{center}
\end{figure}

Table~\ref{tab_param_dyn} shows the parameters of the \textit{windless} model (``NO $\dot{M}$''), that initially was better reproducing the grid of models (before our a priori selection of only models losing mass). We show in Figure~\ref{visib_windless} the visibility fits for this model.  

We notice that there is no significant difference between the wind and the windless visibility profiles in terms of visibility level, in spite of the great difference in the mass-loss rate. This could be related to the similarity in parameter space of the two models: the model producing winds and the windless one have the same $T$, $C/O$, $L$, piston velocity amplitude, and the only different parameters among the two models are surface gravity log($g$) (and hence the ability to produce mass loss) and the mass $M$. This similarity produces very comparable visibility level, but not shape, since the windless model does not produce mass loss. 

The major difference comparing the wind and the windless models fits (Fig.~\ref{fig_interf_voph}, and Fig.~\ref{visib_windless} respectively) is the shape of the visibility spectra, and this is related to the fact that the windless model does not reproduce the dust formation and wind zone. 

\begin{figure}[!htbp]
\begin{center}
\includegraphics[width=0.7\textwidth, bb=83 65 708 538, angle=180]{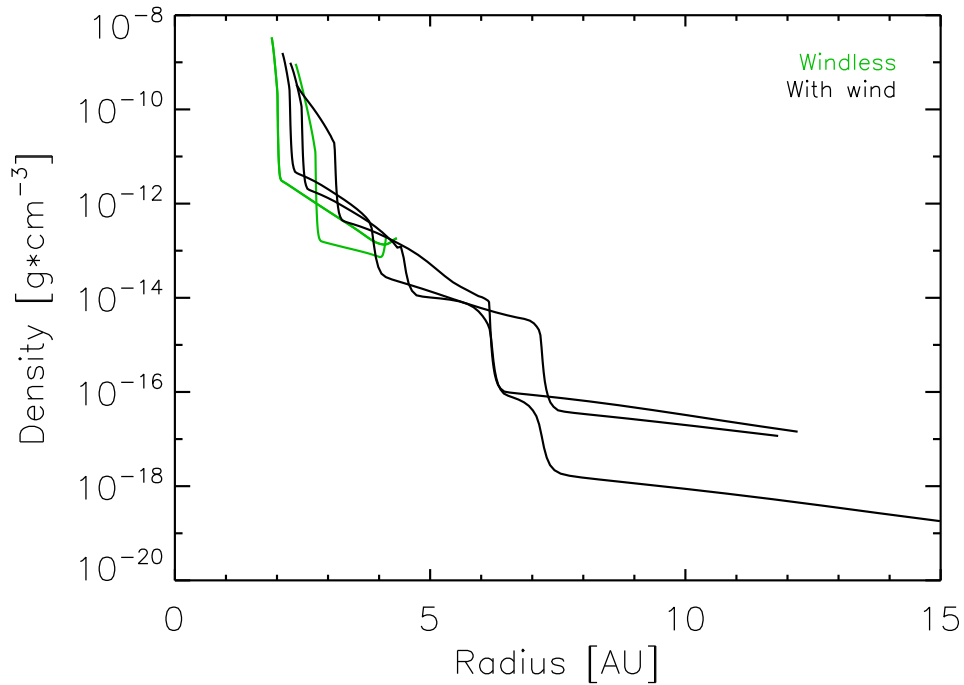}
\caption{Density profiles of the model with wind (black lines), and the windless model (green lines).}
\label{density_overplot}
\end{center}
\end{figure}

Also, comparing the two density profiles (see Fig.~\ref{density_overplot}) the density structure does not differ much within $\sim5$~R$_{\ast}$. This region within $\sim5$~R$_{\ast}$ may be where C$_2$H$_2$+HCN emission and dust emission originate, which primarily affects the MIDI data.

\end{appendices}

\end{document}